\documentclass	[11pt,letterpaper]{article}
\pdfoutput=1
\usepackage	{jheppub}
\usepackage	{journals}
\usepackage	[mathscr]{eucal}
\usepackage	{bm}

\def\None{\mathcal{N}\,{=}\,1}
\def\Ntwo{\mathcal{N}\,{=}\,2}
\def\Nfour{\mathcal{N}\,{=}\,4}
\def\Neight{\mathcal{N}\,{=}\,8}
\def\Nc{N_\text{c}}
\def\Nf{N_\text{f}}
\def\Tc{T_\text{c}}
\def\TC{\mathcal{TC}}
\def\O{\mathcal{O}}
\def\vev#1{\langle#1\rangle}
\def\ket#1{|#1\rangle}

\def\tr{\text{tr}\,}
\def\half{\tfrac{1}{2}}
\def\third{\tfrac{1}{3}}
\def\fourth{\tfrac{1}{4}}
\def\p{\partial}
\def\Atilde{\widetilde{A}}
\def\alphafluc{\widetilde{\alpha}}
\def\chifluc{\widetilde{\chi}}
\def\Htilde{\widetilde{H}}
\def\Ztilde{\widetilde{Z}}
\def\atilde{\widetilde{a}}
\def\ctilde{\widetilde{c}}
\def\vev#1{\langle #1 \rangle}
\def\ket#1{|#1 \rangle}
\def\cs{c_{\rm sound}}
\def\vol{\mathcal V}

\title		{Screening in strongly coupled {\boldmath $\Ntwo^*$}
		supersymmetric Yang-Mills plasma}
\author		{Carlos Hoyos,}
\author		{Steve Paik,}
\author		{Laurence G. Yaffe}
\affiliation	{Department of Physics, University of Washington,
		Seattle, WA 98195, USA}

\emailAdd	{choyos@phys.washington.edu}
\emailAdd	{paik@u.washington.edu}
\emailAdd	{yaffe@phys.washington.edu}

\abstract
    {
    Using gauge-gravity duality, we extend thermodynamic studies and present
    results for thermal screening masses in 
    strongly coupled $\Ntwo^*$ supersymmetric Yang-Mills theory.
    This non-conformal theory is a mass deformation of
    maximally supersymmetric $\Nfour$ gauge theory.
    Results are obtained for the
    entropy density, pressure, specific heat, equation of state, 
    and screening masses,
    down to previously unexplored low temperatures.
    The temperature dependence of screening masses 
    in various symmetry channels,
    which characterize the longest length scales over which thermal 
    fluctuations in the non-Abelian plasma are correlated,
    is examined and found to be asymptotically 
    linear in the low temperature regime.
    }

\arxivnumber	{TBD}
\begin		{document}
\maketitle


\section	{Introduction}


A fundamental characteristic of any system in thermal equilibrium is the
correlation length:
the longest distance scale 
over which spatial fluctuations are significantly correlated.
More precisely, the correlation length describes the slowest
exponential fall-off of correlation functions
at asymptotically large separation.
One may choose to restrict attention to correlators of operators
with specified symmetries,
and define correlation lengths in individual symmetry channels.
Inverse correlation lengths have units of energy
and are commonly referred to as \emph{screening masses}.
These describe the long distance fall-off of correlations in fluctuations
with specified quantum numbers, and hence characterize the nature of
infrared effective degrees of freedom.

In, for example, a relativistic QED plasma,
Debye screening leads to exponential fall-off of
the electric field induced by an external test charge,
$\vev{\vec{E}(\mathbf{x})} \propto e^{-|\mathbf{x}|/\xi}$,
and corresponding behavior
in the correlator of electric field fluctuations.
The inverse correlation length,
or Debye screening mass, $m_{\rm D} \equiv 1/\xi $, is $ O(eT)$.
Fermionic fluctuations in a weakly coupled relativistic plasma have correlations
which decrease exponentially with distance on a shorter
$O(1/T)$ length scale,
corresponding to the inverse of the lowest fermionic Matsubara frequency.
Static magnetic fields are not screened at long distances,
so the associated correlation length is infinite.

In a non-Abelian QCD plasma, one may isolate the physics of
color-electric screening in a gauge-invariant and non-perturbative
manner by examining correlators of operators which are odd under
Euclidean time reflection \cite{ArnYaf}. 
The Debye screening length can be defined as the longest correlation
length in time-reflection odd symmetry channels.%
\footnote{%
Euclidean time reflection $R_\tau$ corresponds to
the product $\TC$ of time reversal and charge conjugation.
This definition of the Debye screening length applies only to 
$\TC$-invariant equilibrium states in $\TC$-invariant theories.
}
At asymptotically high temperatures the Debye screening length
may be calculated perturbatively. 
For an $SU(\Nc)$ gauge group with $\Nf$ Dirac fermions
one finds, at leading order,
$m_{\rm D} = \left[{\frac 13 \Nc+\frac 16\Nf}\right]^{1/2}\,g(T)T$.
Long distance properties of QCD at 
high temperature are effectively described by three-dimensional
pure Yang-Mills gauge theory with a dimensionful coupling
$g_3^2 \equiv g(T)^2 T$.
This theory exhibits three-dimensional confinement and the generation
of a non-perturbative $O(g_3^2)$ mass gap.
Therefore, the longest distance correlations in asymptotically
high temperature QCD are associated with static magnetic fluctuations,
and have an $O[(g(T)^2 T)^{-1}]$ correlation length.

At non-asymptotic temperatures,
reliable calculations of QCD screening masses require numerical
lattice simulations.%
\footnote
    {%
    For example,
    up to temperatures as high as $10^7 \,\Tc$,
    the non-perturbative $O(g^2 T)$ correction to the
    Debye screening mass in QCD is larger than
    the leading $O(gT)$ perturbative contribution
    \cite{Laine}. 
    }
At temperatures of a few times the deconfinement temperature,
$T/\Tc \sim 1$--4,
which is the temperature range probed by heavy ion collisions,
the dynamics of QCD plasma is strongly coupled.
Evidence for the strongly coupled nature of QCD plasma in this
regime includes the results for screening masses obtained from
lattice simulations (see ref.~\cite{Laer} for a review),
as well as the success of low viscosity hydrodynamic simulations
in reproducing the collective flow observed at RHIC,
and the observed suppression of high transverse momentum jets in
RHIC collisions \cite{Tannen}.

Numerical lattice simulations are limited to equilibrium
Euclidean observables; with very limited exceptions,
non-equilibrium Minkowski space dynamics cannot be
extracted reliably from lattice simulations.
Consequently,
there has been much interest in studying
dynamic properties of non-Abelian gauge theories
which mimic some aspects of QCD and to which the
techniques of gauge/gravity (or AdS/CFT) duality \cite{adscftA,adscftB,adscftC}
may be applied.
(See, for example, refs.~\cite{shear1,shear2,boost,fluidgrav,drag1,drag2,drag3,quench,thermalization}.)
Gauge/gravity duality reformulates the strong coupling, large $\Nc$ 
dynamics of a suitable gauge theory in terms of classical supergravity
in an asymptotically anti-de Sitter spacetime.
The simplest, and best understood, example is maximally
supersymmetric $SU(\Nc)$ Yang-Mills theory ($\Nfour$ SYM).
Although this theory contains adjoint representation fermions
and scalars not present in QCD, at non-zero temperature the
resulting non-Abelian plasma shares many qualitative similarities
with the quark-gluon plasma of QCD.
Screening masses obtained via 
gauge-gravity duality for strongly coupled ($\lambda \equiv g^2 \Nc \gg 1$)
$\Nfour$ SYM,
in the $\Nc \to \infty$ limit,
were discussed in refs.~\cite{Bak, Amado}.
The conformal invariance of $\Nfour$ SYM implies that thermal screening
masses are exactly proportional to $T$. 
Ref.~\cite{Bak} compared the values of
screening masses of strongly coupled $\Nfour$ SYM
with results of lattice QCD simulations
at $T \approx 2\Tc$.
In QCD, this temperature
lies within the window $1.5\,\Tc \leq T \lesssim 4\,\Tc$ where 
screening masses are well described as scaling linearly with $T$ \cite{Datta2,DG}.
Ratios of screening masses in different symmetry channels
compare rather well between QCD and $\Nfour$ SYM
but, in absolute size, the $\Nfour$ SYM screening masses (divided by $T$)
are roughly a factor of two larger than in QCD \cite{Bak}.

It is not currently known whether quantitative differences
between QCD and $\Nfour$ SYM plasma properties
are dominantly due to
(\emph{i}) the differing constituents of the non-Abelian plasmas,
(\emph{ii}) the comparison of QCD at $T/\Tc \approx 1.5$--4, where it is
neither weakly coupled nor infinitely strongly coupled,
with the $\lambda \gg  1$ limit of SYM instead of some $O(1)$ value
of $\lambda$, or
(\emph{iii}) the comparison with $\Nc = \infty$ SYM instead of $\Nc = 3$.
Available lattice results suggest that the $\Nc$ dependence of
plasma thermodynamics and screening masses
is quite mild \cite{Datta2,Hart}, so possibility (\emph{iii})
is unlikely to be dominant.
Possibility (\emph{ii}) could be studied by computing $1/\lambda$
corrections to the leading strong coupling behavior described by
classical supergravity.
However, for many observables this requires knowledge of higher derivative
corrections to the supergravity action which have not yet been
fully worked out.
The possibility most accessible to study is (\emph{i}):
the dependence of various observables on the particular constituents
of a non-Abelian plasma.
Part of the reason that screening masses are larger in
$\Nfour$ SYM than in QCD
is certainly the fact that there are more active degrees of freedom
contributing to screening in an $\Nfour$ SYM plasma --- adjoint
representation fermions
and scalars instead of fundamental representation quarks.
One would expect that reducing the number of light degrees of freedom
would yield screening masses which are closer to those of a QCD plasma.
This may be explored, in strongly coupled plasmas,
by studying theories with known gravity duals which are
more QCD-like than $\Nfour$ SYM.
Examining non-conformal deformations of $\Nfour$ SYM will
shed light on the sensitivity of ratios of screening masses
to departures from conformal invariance.

In this paper,
we explore these issues by computing
screening masses and thermodynamics
of a mass-deformation of maximally supersymmetric Yang-Mills
theory known as $\Ntwo^*$ SYM \cite{Donagi,PW,BLthermo}.
This theory differs from $\Nfour$ SYM by the addition
of mass terms and interactions for an adjoint representation hypermultiplet,
in a manner which preserves $\Ntwo$ supersymmetry.
This theory has one new free parameter, the deformation
mass $m$.
The theory is non-conformal, with a nontrivial renormalization
group fixed point (with adjustable coupling $\lambda$)
at asymptotically high energies.
An intrinsic strong scale $\Lambda$ appears,
analogous to $\Lambda_\text{QCD}$,
but for large $\lambda$ this scale is comparable
to the mass deformation, $\Lambda \sim m$ \cite{Donagi}.

The study of $\Ntwo^*$ gauge theory at non-zero temperature using
holographic methods was pioneered by Buchel, Liu, and collaborators
\cite{BLthermo, Bhydro,BDKL}.
Various properties of $\Ntwo^*$ plasma have been studied,
including shear and bulk viscosities
\cite{Bhydro, Bviscosity, Bviscosity2}, as well as
heavy quark drag, momentum broadening, and jet quenching \cite{Hoyos}.
Reference~\cite{BDKL} laid the groundwork for numerically constructing
the supergravity background dual to the thermal equilibrium state
of $\Ntwo^*$ SYM, and mapped out the dictionary that
translates between supergravity fields and observables in the quantum
field theory. Of direct relevance to our work is their result for
the free energy density in the range $0 \leq m/T \leq 6$.%
\footnote
    {%
    In the later work \cite{Bviscosity}, hydrodynamic coefficients
    were evaluated up to $m/T \simeq 12$.
    }
Noticeable
deviation from the behavior of $\Nfour$ plasma was estimated to
occur near the upper end of this range.

We extend the results of ref.~\cite{BDKL}  
on $\Ntwo^*$ thermodynamics to substantially lower temperatures,
$m/T \lesssim 33$, and provide evidence that this data allows one
to probe the asymptotically low temperature regime.
We argue that, despite the presence of a non-zero mass scale $m$,
this theory (in the large $\Nc$ and large $\lambda$ limit)
has no thermal phase transition at any non-zero temperature.

Our primary goal is the evaluation of
thermal screening masses for multiple
symmetry channels throughout the temperature range $0 \le m/T \lesssim 33$.
We extrapolate the screening mass curves to vanishingly small
temperatures, $T/m \to 0$, where components of the heavy
hypermultiplet decouple from the dynamics due to exponential Boltzmann
suppression, and extract results for the limiting
plasma which contains only light degrees of freedom.

The paper is structured as follows. In Sec.~\ref{sec:review},
we briefly review the field theoretic formulation of $\Ntwo^*$ gauge
theory and its gravitational representation.
In Sec.~\ref{sec:thermo},
we discuss equilibrium thermodynamic quantities such as the entropy density,
pressure, specific heat, and equation of state.
We compute screening masses from linearized
fluctuations of supergravity modes in Sec.~\ref{sec:masses}.
In Sec.~\ref{sec:compare},
we compare the screening masses of $\Ntwo^*$ SYM to those of $\Nfour$
SYM and to available lattice QCD data.
The holographic renormalization and derivation of thermodynamic quantities
from the gravitational dual are discussed, at some length,
in Appendix~\ref{app:background}.
The numerical method we used to find the background geometry is explained
in Appendix~\ref{app:numerics}, and fluctuations of the gravitational fields
we use to compute screening masses are described in Appendix \ref{app:flucs}.
Additional material related to the appendices can be found online
\cite{online}.


\section	{$\Ntwo^*$ gauge theory and its gravity dual}
\label		{sec:review}


$\Ntwo^*$ supersymmetric Yang-Mills theory
is a mass deformation of $\Nfour$ SYM. We consider the
theory on $\mathbb{R}^{1,3}$ with gauge group $SU(\Nc)$ in the 't Hooft limit: $\Nc \to \infty$ and
$g^2 \to 0$ with $\lambda \equiv g^2\Nc$ held fixed. 
If the $\Nfour$ field content is grouped into $\None$ superfields,
there is a vector multiplet and three adjoint chiral multiplets $\Phi_i$
($i=1,2,3$).
The Lagrange density of $\Ntwo^*$ SYM is obtained by adding
the term
\begin{equation}
\delta\mathcal{L} = m\int d^2\theta\> \tr(\Phi_1\Phi_2) + \text{h.c.}
\end{equation}
to the $\Nfour$ SYM Lagrange density $\mathcal{L}_{\Nfour}$.
Appropriate field redefinitions can remove any phase in the mass,
so $m$ may be taken real and positive without loss of generality. 
The two chiral fields $(\Phi_1, \Phi_2)$ comprise a single
$\Ntwo$ massive hypermultiplet,
and the remaining fields form an $\Ntwo$ vector multiplet.
Once auxiliary field constraints
have been solved, the above contribution to the
superpotential induces conventional mass terms for two Weyl fermions
and two complex scalars, 
as well as terms trilinear in the scalars. In order to identify the gravity
dual of the $\Ntwo^*$ SYM theory, it is convenient to write the mass deformation in 
terms of relevant operators in irreducible representations of the $\Nfour$ $R$-symmetry
group $SO(6)_R$. We can distinguish two contributions, first there is an irreducible scalar mass
\begin{equation}\label{eq:operatorO2}
\O_2 \equiv \third\tr\bigl(-|\phi_1|^2 - |\phi_2|^2 + 2|\phi_3|^2\bigr).
\end{equation}
This operator, which is an element of the $\mathbf{20'}$ of $SO(6)$,
appears in the 4D (Euclidean) $\Ntwo$* action
$S_{\Ntwo*}$ as $-(2/g^2)\int d^4x\, m^2\, \O_2$.
Hence, it
contributes the usual positive bosonic mass terms for the hypermultiplet
scalars $\phi_1$ and $\phi_2$.
But it also destabilizes the scalar $\phi_3$ belonging to the vector multiplet.
The second contribution introduces a mass for the fermions in the hypermultiplet
\begin{equation}\label{eq:operatorO3}
\O_3 \equiv \tr\bigl(i\psi_1\psi_2
- \sqrt{2}\phi_3[\phi_1, \phi_1^\dag] - \sqrt{2}\phi_3[\phi_2^\dag, \phi_2] + \text{h.c.}\bigr)
- \tfrac{2}{3}\, m\,\tr\bigl(|\phi_1|^2 +|\phi_2|^2 +|\phi_3|^2\bigr),
\end{equation}
and appears in $S_{\Ntwo*}$ as $-(2/g^2)\int d^4x\> m\,\O_3$. One may obtain
the expression for $\O_3$ by first deriving the on-shell Lagrange density for
mass-deformed $\Nfour$ SYM in terms of component fields, isolating the
$m$-dependent terms, and subtracting 
$m^2 \, \O_2$ from them.
Note that $\O_3$ is modified, from its massless limit,
by the $SO(6)$ singlet operator
$\sum_1^3 |\phi_i|^2$
multiplied by a factor of $m$. This term is
crucial since it cancels the negative potential energy for $\phi_3$ introduced by $\O_2$. 
It is required by the central extension of the $\Ntwo$ supersymmetry algebra.%
\footnote
{%
To see this, 
consider the identity $\{Q_\alpha^A, Q_\beta^B\} = \epsilon_{\alpha\beta} \, Z^{AB}$. To obtain a 
massive multiplet which has 4 helicity states (2 integer and 2 odd half-integer), the Bogolmolny
bound must be saturated: $Z^{12} = -Z^{21} = 2m$. This value for the antisymmetric matrix $Z^{AB}$
forces half of the $2\mathcal{N}$ fermionic ``oscillators" (obtained from suitable linear 
combinations of the supercharges) to obey trivial anticommutation relations. This means that 
they cannot be
used to create or annihilate additional helicity states. This, in turn, shortens a massive
irreducible representation of the $\mathcal{N}$-extended super-Poincar\'e algebra from dimension
$2^{2\mathcal{N}}$ to $2^\mathcal{N}$, which is precisely what is needed to have a massive
hypermultiplet. The action of $Q^1$ and $Q^2$ applied successively to $\O_2$ must yield the 
bosonic operator $\O_3$, and the anticommutation relation implies that an explicit factor of $m$
must appear in $\O_3$.

The dual operator analogous to $\O_3$ in the case of fundamental representation 
hypermultiplets is given explicitly in Appendix A of ref.~\cite{Myers}. Specifically,
in eq.~(A.1) of this reference,
the operator $\O_\text{m}$ has an identical fermion bilinear, similar 
scalar trilinears, and similar mass terms for the hypermultiplet scalars. However, unlike $\O_3$, 
commutators do not appear in the scalar trilinears of $\O_\text{m}$. The reason is that 
their superpotential (in our notation) is $\tr(\Phi_1\Phi_3\Phi_2)$ rather than 
$\tr(\Phi_3[\Phi_1,\Phi_2])$. We thank Andreas Karch for pointing this out to us.
}

Since the UV fixed point describes the conformal $\Nfour$ theory,
whose coupling is a free parameter,
one has the freedom to choose the UV coupling
$g^2$ (or $\lambda$) arbitrarily.%
\footnote{%
One may also choose an arbitrary $\theta$ angle,
but this will not be important in our discussion.
}
We choose the 't Hooft coupling to be large, $\lambda \gg 1$,
but finite as $\Nc \to \infty$.%
\footnote
    {%
    This is necessary since
    the supergravity approximation to the 
    string partition function is only valid for spacetime curvatures
    which are small when measured in string units.
    In our case, the ratio $\ell_\text{s}/L$
    of the fundamental string length $\ell_\text{s}$
    to the characteristic curvature radius $L$
    equals $\lambda^{-1/4}$.
    }
The value of this coupling, together with the mass $m$,
determines the intrinsic scale where the theory becomes strongly coupled,
$\Lambda \sim m \, e^{-4\pi^2/\lambda}$ \cite{Donagi}. 
Notice, however, that for strong coupling the scale $\Lambda$
is comparable to $m$ and there is no large separation of scales.
Consequently, one cannot decouple massive modes at the scale $m$ 
while preserving the low energy spectrum at the scale $\Lambda$.

The global symmetry group is $SU(2)_R \times U(1)$.
The $U(1)$ is an ordinary flavor symmetry
under which $\Phi_1$ has charge $+1$, $\Phi_2$ has charge $-1$,
and $\Phi_3$ is neutral. 
The superpotential term $m\, \tr(\Phi_1\Phi_2)$
does not preserve the classical $U(1)_R$ symmetry
that exists in the massless $\Nfour$ theory.
However, one can regard $m$ as the expectation value of a background field
and assign it an $R$-charge of 2.
This implies that the low energy effective action must be holomorphic in $m$.
The vacuum manifold
is an $(\Nc{-}1)$-dimensional complex space specified by adjoint scalar 
expectation values having the form
$\vev{\Phi_1} = \vev{\Phi_2} = 0$ and
$\vev{\Phi_3} = \text{diag}(a_1, \dotsc, a_{\Nc})$
(up to a gauge transformation),
subject to the tracelessness constraint, $\sum_i a_i = 0$.
At a generic point in moduli space,
$\Ntwo$ supersymmetry is unbroken, the gauge group is Higgsed to
$U(1)^{\Nc-1}$, and the phase is Coulombic.
A Wilsonian effective action for the low-energy Abelian theory may be
generated by integrating out massive $W$ bosons, their superpartners,
and the adjoint hypermultiplet. 
Two derivative terms
in the effective action are determined from a prepotential which depends
holomorphically on the complexified gauge coupling constant
$\tau \equiv \theta/(2\pi) + 4\pi i/g^2$, the mass $m$, 
and the eigenvalues $\{a_i\}$ \cite{BPP}. The prepotential
receives classical and quantum contributions.
The perturbative correction is one-loop exact and may
be determined by a matching calculation.
Nonperturbative corrections become important in regions
of moduli space where BPS states become light.
Both monopole and $W$ boson masses are proportional to
differences of scalar eigenvalues, $|a_i{-}a_j|$,
and vanish when eigenvalues coincide.
Away from points of eigenvalue degeneracy,
nonperturbative corrections are suppressed and the
low energy dynamics is well-described by
the perturbative prepotential.

In gauge/gravity duality, certain gravitational backgrounds
may be interpreted as dual descriptions of $\Nfour$ SYM perturbed
by relevant operators. In particular, solutions of 5D
maximally supersymmetric $\Neight$ gauged supergravity which asymptotically approach $AdS_5$ are
dual to states of $\Nfour$ SYM with gauge group $SU(\Nc)$.
If fields in addition to the metric are non-vanishing,
then one can obtain non-conformal boundary theories with non-trivial
renormalization group flow.
Because 5D gauged supergravity is a consistent truncation 
of 10D type IIB supergravity on $AdS_5 \times S^5$,
it is sufficient to solve the dimensionally-reduced problem
(as opposed to the full 10D problem) \cite{sugra1,sugra2,sugra3}.
It is possible to unambiguously ``uplift" 
any 5D solution to a full 10D solution although,
in practice, finding the uplift is nontrivial. 
The scalar and gravity part of the 5D action has the form
\begin{equation}
\label{action}
S_{5D} = \frac{1}{4\pi G_5}\int d^5x\> \sqrt{-g}\,
\left(\fourth \, R + \mathcal{L}_\text{matter}\right).
\end{equation}
Specific terms in $\mathcal{L}_\text{matter}$ will be discussed below.
Massless bosonic modes in
10D may be expanded in a set of Kaluza-Klein modes on the $S^5$,
with masses of order $1/L$, where $L$ is the radius of the five-sphere.
There are 42 scalar fields and a complicated potential $V$ 
which depends on 40 of them \cite{sugra1,sugra2,sugra3}. For a thorough description of the subsector we will study here, see refs.~\cite{PW,KPW}.
The scalars fall into various representations of the $SO(6)$ gauge group:
there is a $\mathbf{20'}$ representation with mass-squared 
$M^2 = -4/L^2$, a $\mathbf{10} \oplus \overline{\mathbf{10}}$
representation with $M^2 = -3/L^2$,
and dilaton and axion singlets which have $M^2 = 0$ and do not
contribute any potential energy.
The scaling dimension $\Delta$ of a CFT operator $\O_\Delta$ 
dual to a supergravity scalar field 
with mass-squared $M^2$ on $AdS_5$ is given by $\Delta(\Delta-4) = M^2 L^2$.
The $SO(6)$ symmetry maps to the global $R$-symmetry in the $\Nfour$ SYM theory.
The $\mathbf{20'}$ is dual to dimension~2 symmetric traceless combinations of
the six $\Nfour$ scalars (i.e.,
the real and imaginary parts of $\phi_i$), the $\mathbf{10} \oplus \overline{\mathbf{10}}$ 
is dual to dimension 3 symmetric bilinears of the four $\Nfour$ Weyl fermions plus 
supersymmetric completions, and the dilaton-axion pair is dual to the $\Nfour$ Lagrange density 
plus theta term.%
\footnote{%
The boundary value of the 5D dilaton-axion pair is the UV marginal coupling $\tau$. The physical
running coupling, dual to the 10D dilaton and axion, is a nontrivial function of $\tau$, $m$, and
the energy scale \cite{PW}.
} 
The scalar fields have $M^2 \leq 0$
(but above the Breitenlohner-Freedman stability bound),
which corresponds to $\Delta \leq 4$. Therefore,  
a solution of 5D $\Neight$ gauged supergravity with non-vanishing
profiles for any of the 40 nontrivial 
scalars corresponds to a relevant deformation of the boundary CFT.

In our case, the deformations are encoded in two real supergravity scalars $\alpha$ and $\chi$,
in terms of which the 
5D matter Lagrange density equals%
\begin{align}
-\mathcal{L}_\text{matter} &= 3(\p\alpha)^2 + (\p\chi)^2 + V(\alpha,\chi),
\label{eq:Lmat}
\\
\noalign{\noindent with}
V(\alpha,\chi) &= \hat{g}^2\Bigl[-\fourth e^{-4\alpha} - \half e^{2\alpha} \cosh 2\chi + 
\tfrac{1}{16}\, e^{8\alpha} \sinh^2 2\chi\Bigr].
\label{eq:Vscalar}
\end{align}
The dimensionful gauged supergravity coupling is $\hat{g}^2 = (2/L)^2$, and the 5D
Newton's constant is $G_5 = \pi L^3/(2\Nc^2)$.
The field $\alpha$ has $M^2 = -4/L^2$ and is dual to the $\Ntwo$* operator $\O_2$ in eq.~\eqref{eq:operatorO2}. 
The second scalar field $\chi$ has $M^2 = -3/L^2$ and is dual to the $\Ntwo^*$ 
operator $\O_3$ in eq.~\eqref{eq:operatorO3}.
The value of the mass $m$ in the field theory is determined by the asymptotic boundary conditions of the fields $\alpha$ and 
$\chi$, which are not independent but related by supersymmetry. Notice that the identification of $\chi$ with $\O_3$ implies that
the map between the field and the dual operator depends on the choice of boundary conditions/couplings in the field theory Lagrangian.
 This modification of the usual prescription may be a consequence of using a Kaluza-Klein compactification of ten-dimensional supergravity, 
we consider preserving supersymmetry on both sides as an indication that the identification is correct.

A solution to 5D $\Neight$ gauged supergravity with non-vanishing scalars
$\alpha$ and $\chi$ is known, 
and its full 10D uplift has been constructed \cite{PW}. The bulk 5D geometry has four-dimensional 
Poincar\'e invariance. An analysis using a slowly-moving probe D3-brane indicates
that the background branes form a locus around the origin of the probe's
moduli space (the space where the probe's potential vanishes) where the probe's kinetic energy 
vanishes. This `enhan\c{c}on' geometry 
is dual to a special vacuum of the $\Ntwo^*$ Coulomb branch where extra dyonic states become 
massless \cite{BPP, Evans}. We will see that this solution plays an important role in the low
temperature regime.


\section	{Equilibrium thermodynamics}
\label		{sec:thermo}


Buchel, Deakin, Kerner, and Liu numerically solved the 5D $\Neight$
gauged supergravity equations for a background with
three-dimensional rotational invariance and a regular black brane horizon 
\cite{BDKL}.%
\footnote{%
The zero temperature solutions were previously constructed analytically by Pilch and Warner \cite{PW}.
} 
They also verified that the 10D uplift solves the IIB 
supergravity equations of motion. This geometry is dual to a thermal equilibrium state of 
strongly coupled $\Ntwo^*$ SYM with gauge group $SU(\Nc)$ in the large $\Nc$ limit.
There is a single dimensionless parameter,
$m/T$, which may be freely adjusted.
Thermodynamic quantities,
when divided by appropriate
powers of $T$, must be dimensionless functions of $m/T$.

We have extended the analysis of ref.~\cite{BDKL} to a wider range of $m/T$.
The holographic representation of the entropy and free energy
is summarized in Appendix~\ref{app:background}.
As $m/T$ increases, the differential equations one must solve become increasingly stiff,
making the numerical exploration of very large values of $m/T$ challenging.

\begin{figure}[t]
    \centerline{
	\includegraphics[scale=0.88]{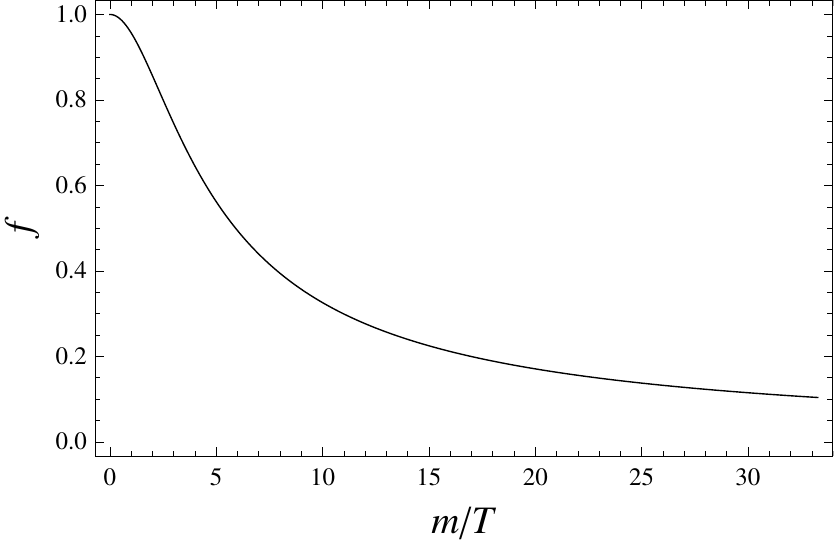}\hfill
	\includegraphics[scale=0.88]{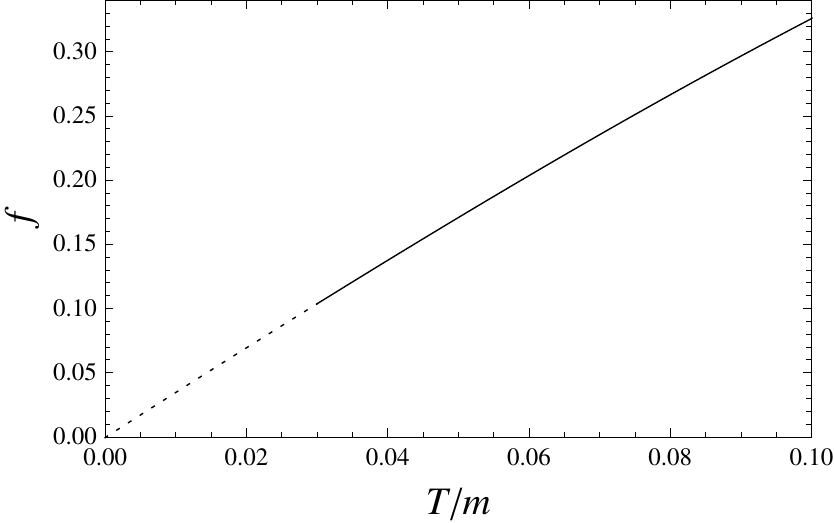}
	}
    \vspace*{-10pt}
    \caption{Left: Pressure
    divided by $\tfrac{\pi^2}{8}\Nc^2 T^4$,
    as a function of $m/T$.
    Right: Low temperature behavior plotted as a function of $T/m$.
    The dotted line shows the extrapolation of a quadratic fit to the ten
    lowest-temperature data points.}
    \label{fig:free}
\end{figure}

The pressure (equal to minus the free energy density) may be expressed in the form
\begin{equation}
\label{free_energy_gauge}
p = \frac{\pi^2}{8}\, \Nc^2 T^4 f(m/T) \,.
\end{equation}
The scaling function $f$ is plotted in figure~\ref{fig:free}. 
The free energy is obtained from the renormalized Euclidean supergravity action
(see Appendix~\ref{app:holorg}).
The prefactors in eq.~(\ref{free_energy_gauge})
have been chosen to equal the $\Nfour$ value of the pressure,
so that $f(0) = 1$.
A high temperature expansion yields
$f = 1 - 2\pi^{-4}\, \Gamma(\tfrac 34)^4\,(m/T)^2 + O[(m/T)^3]$ \cite{BLthermo}.
Our numerical 
computations extend out to $m/T \approx 33.3$, at which point $f \approx 0.1037$. 
Throughout the range of temperatures $0 \leq m/T \lesssim 33$,
our numerically-determined black brane solution describes a thermodynamic state
which is at least locally stable.
Both the pressure and the specific heat (discussed below) are positive,
and we presume that our solution is describing the true equilibrium state.
As shown in the right-hand plot of figure~\ref{fig:free},
extrapolation of our numerical results clearly suggest that
$f$ vanishes linearly as $T/m \to 0$,
implying that the free energy density $F/\vol = O(T^5/m)$ at low temperature.
The interpretation of this scaling behavior is discussed below.

\begin{figure}[t]
    \centerline{\includegraphics[scale=1.0]{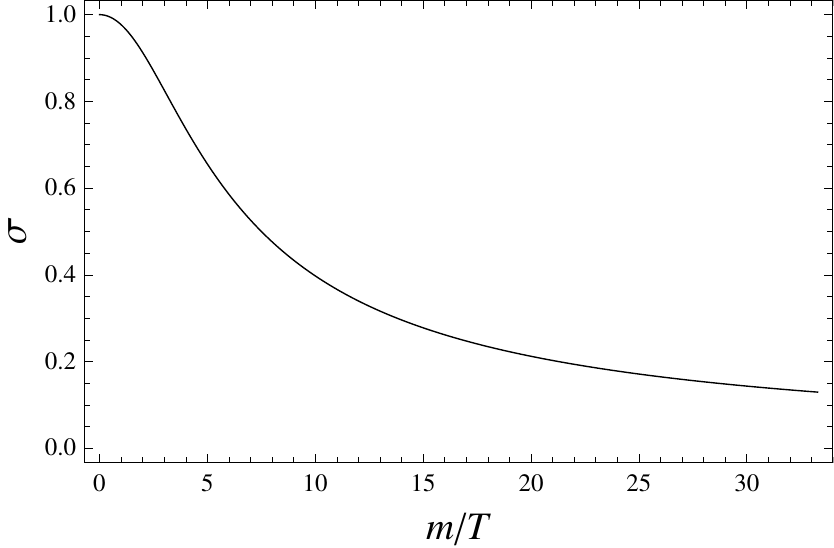}}
    \vspace*{-10pt}
    \caption{The entropy density,
    divided by $\tfrac{\pi^2}{2}\,\Nc^2 T^3$, as a function of $m/T$.
    }
    \label{fig:entropy}
\end{figure}

The entropy density may be expressed as
\begin{equation}
\label{entropy}
S/\vol = \frac {\pi^2}{2}\, \Nc^2 T^3\, \sigma(m/T),
\end{equation}
where the prefactor equals the entropy density of $\Nfour$ SYM \cite{Malda},
so that $\sigma(0) = 1$.
The entropy may be obtained from the horizon area \eqref{dict:entropy}
or from a thermodynamic derivative of the free energy,
$S = -\partial F/\partial T$,
which implies that $\sigma = f - \frac 14 \, \frac mT \, f'$.
The resulting scaling function $\sigma$ is plotted in figure~\ref{fig:entropy}.
For $m/T \ll 1$, a high temperature expansion yields
$\sigma = 1 - \pi^{-4}\, \Gamma(\tfrac 34)^4 \, (m/T)^2 + O[(m/T)^3]$ \cite{BLthermo}.
As the temperature drops, $\sigma$ decreases
and appears to approach zero linearly in $T/m$,
implying that the entropy density $S/\vol = O(T^4/m)$ at low temperature.

\begin{figure}[t]
    \centerline{\includegraphics[scale=1.0]{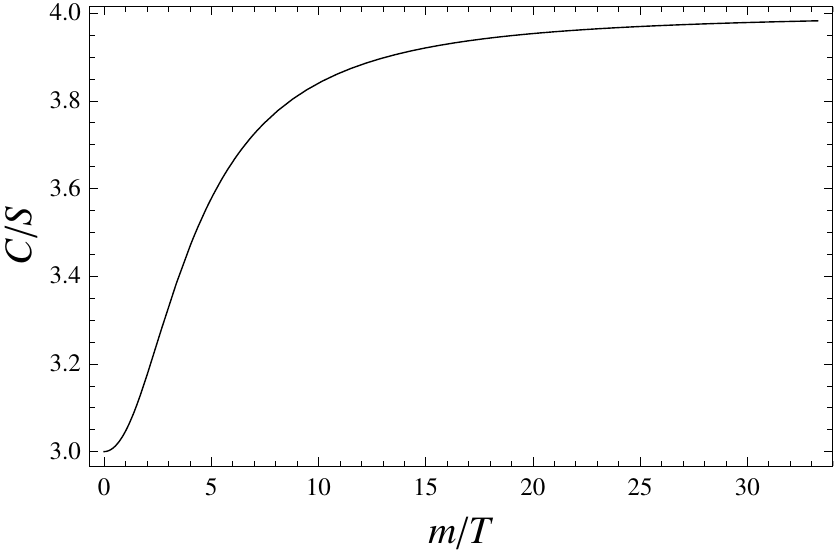}}
    \vspace*{-10pt}
    \caption{The specific heat to entropy ratio, $C_V/S = \p(\ln S)/\p(\ln T)$,
    as a function of $m/T$.}
    \label{fig:specificheat}
\end{figure}

The specific heat at constant volume is given by
\begin{equation}
C_V = T\left(\frac{\p S}{\p T}\right)_V
    = S\left(3 - \frac{m}{T}\,\frac{\sigma'}{\sigma}\right) .
\end{equation}
A plot of the specific heat to entropy ratio, $C_V/S$,
is shown in figure~\ref{fig:specificheat}, from which
it is evident that the specific heat is always positive. 
The speed of sound (squared) in the plasma is inversely related to the
specific heat,
$\cs^{-2} = C_V/S$.
Figure~\ref{fig:specificheat} shows that $\cs^2$ equals
$1/3$ in the high temperature (or massless) limit, as required from conformal invariance,
with corrections quadratic in $m$.
Explicitly,
$
    \cs^2 = \frac{1}{3}
    - \frac{2}{9}\pi^{-4}\,{\Gamma(\frac{3}{4})^4}\, (m/T)^2 + O[(m/T)^3]
$.
At low temperatures, the speed of sound (squared) evidently approaches $1/4$.

\begin{figure}[t]
    \centerline{\includegraphics[scale=1.0]{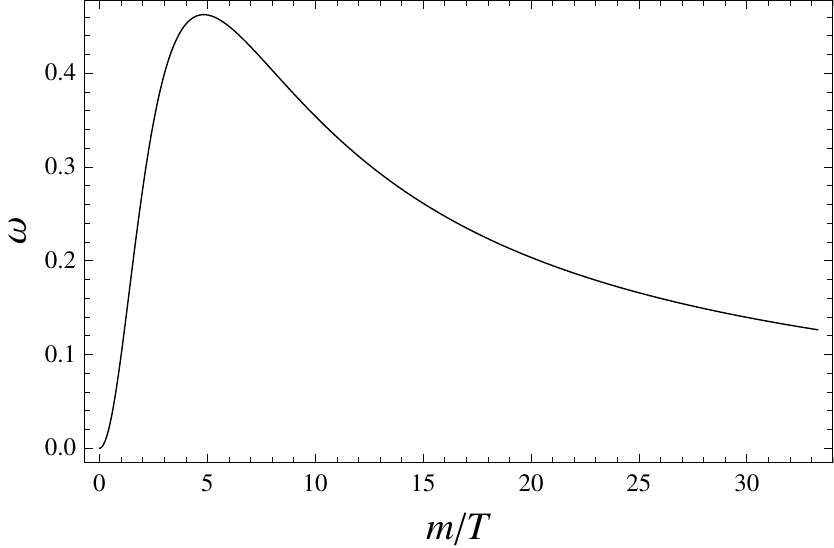}}
    \vspace*{-10pt}
    \caption{The scaling function for the equation of state,
             $\omega = (\epsilon - 3p)/(\Nc^2 T^4)$,
	     as a function of $m/T$.}
    \label{fig:trace}
\end{figure}

The equation of state, relating the energy density $\epsilon$ and pressure $p$ to the
temperature, may be expressed as
\begin{equation}
\label{EoS}
\epsilon-3p = \Nc^2 T^4 \, \omega(m/T),
\end{equation}
where
\begin{equation}
\omega
    = \frac {\pi^2}{2} \, (\sigma -f)
    = -\frac {\pi^2}{8} \, \frac mT \, f' \,.
\end{equation}
This function is plotted in figure~\ref{fig:trace}.
Since $\epsilon-3p$ is the trace of the
energy-momentum tensor,
it vanishes in the conformal limit.
For small $m/T$ the scaling function $\omega$ rises quadratically,
$\omega = \tfrac 12 \pi^{-2}\,\Gamma(\frac 34)^4 (m/T)^2 + O[(m/T)^3]$ \cite{BLthermo}.
As shown in figure~\ref{fig:trace},
$\omega$ peaks at $m/T \approx 4.83$ and
decreases monotonically thereafter.
The position of the peak corresponds to the temperature below which the
plasma is cool enough that massive components of the hypermultiplet have a low probability 
of being thermally excited.
The high-mass/low-temperature tail of $\omega$ must be predominately due
to the non-renormalizable ({i.e.}, higher-derivative) interactions
of the light degrees of freedom.
The fact that $m/T \approx 33$ lies well out in the tail of $\omega$, far from the peak,
is evidence that our numerical results are probing behavior deep into
the low temperature regime.

\begin{figure}[t]
    \centerline{\includegraphics[scale=0.61]{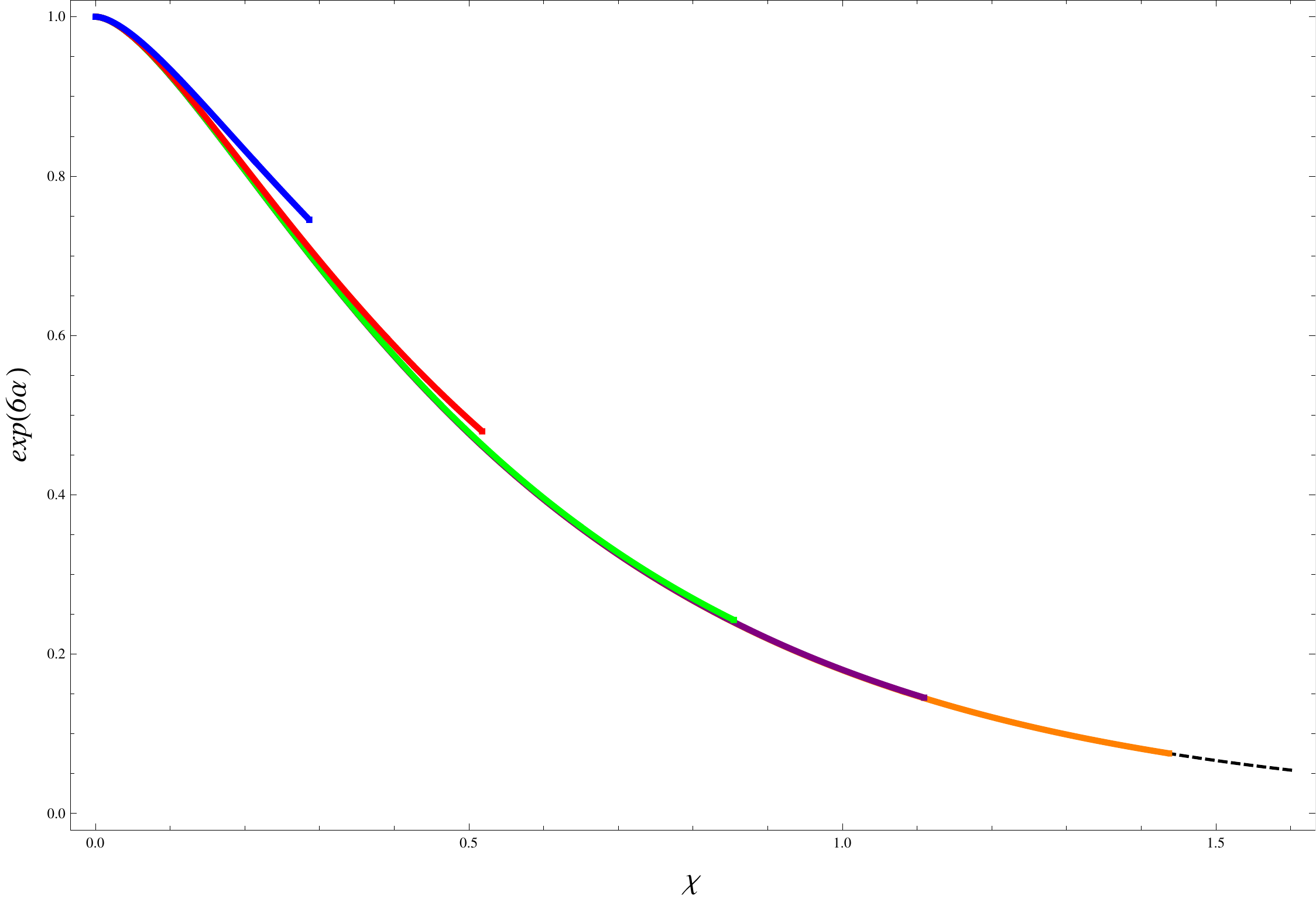}}
    \vspace*{-10pt}
    \caption{Evolution of the scalar fields $(\chi, e^{6\alpha})$ as the
    radial coordinate varies. All curves start at $(0, 1)$, corresponding to
    the boundary where the scalars $\alpha$ and $\chi$ vanish. Curves end
    at the point $(\chi_h, e^{6\alpha_h})$ corresponding
    to the value of the scalars at the horizon.
    The $T=0$ enhan\c{c}on solution is shown as a dashed curve extending from $(0, 1)$ to $(\infty, 0)$. 
    Finite temperature solutions are shown 
    as solid lines for $m/T \approx 2.23$ (blue), $4.57$ (red), $10.0$ (green), 
    $17.1$ (purple), and $33.3$ (orange).
    Our finite temperature solutions clearly approach the
    enhan\c{c}on solution as $m/T \to \infty$.}
    \label{fig:flows}
\end{figure}

The low temperature regime is dominated by the special enhan\c{c}on
solution mentioned in Sec.~\ref{sec:review}.
This is shown by the plot in figure~\ref{fig:flows},
which displays the evolution of scalars fields $(\chi, e^{6\alpha})$
along the radial direction for solutions at different values of $m/T$,
superimposed on the $T=0$ enhan\c{c}on solution,
whose explicit form can be found in Appendix~\ref{app:background}. As the
temperature drops to zero, our solutions clearly approach the enhan\c{c}on
solution. As discussed in ref.~\cite{Carlos}, the dual to the enhan\c{c}on
geometry may be interpreted as a flow from the $\Nfour$ SYM theory
in the UV to a five-dimensional CFT in the IR. At low temperatures
the effective five-dimensional theory dominates the dynamics, so this
explains the behavior of the free energy, $F/\vol \sim T^5/m$, the entropy,
$S/\vol \sim T^4/m$, and the speed of sound, $\cs^2 \to 1/4$, that we observe.%
%
\footnote
    {%
    This is also consistent with
    the result of refs.~\cite{Bviscosity, Bviscosity2}
    that in the zero temperature limit the bulk viscosity $\zeta$ saturates the bound 
    $\zeta/\eta \geq 2(\frac 13 - c_s^2)$, in agreement with ref.~\cite{noncftbranes}.
    }

This low-temperature behavior is presumably specific to the $\Nc \to \infty$ limit.
Turning on an infinitesimal, non-zero temperature will deform the moduli space
of degenerate vacuum states of $\Ntwo$* theory into some non-trivial free-energy surface.
One expects the free energy to be minimized at those points in moduli space
with the maximal number of massless degrees of freedom.
In $\Nfour$ SYM, a non-zero temperature destabilizes all points on moduli space
other than the origin, where all scalar expectation values vanish and the entire
$SU(\Nc)$ gauge group is ``unbroken.''
Conformal invariance guarantees that the non-Abelian plasma phase,
with $O(\Nc^2)$ free energy, extends all the way down to $T = 0$.
For the non-conformal $\Ntwo$ SYM theory, the low temperature behavior is more subtle.
There is no point on the $\Ntwo$ moduli space where a non-Abelian gauge field
is present in the low energy dynamics.
Low energy degrees of freedom consist of $U(1)^{\Nc-1}$ gauge fields plus,
at special points in moduli space, massless BPS monopoles or dyons \cite{SW1,SW2,Argyres}.
One would expect low temperature equilibrium states to correspond to
points where a maximal number of BPS dyons simultaneously become massless.
This suggests that for sufficiently low temperatures there will be
multiple degenerate equilibrium states, related by $SL(2,\mathbb{Z})$ electro-magnetic duality \cite{SW2}.%
\footnote
    {%
    For $\Ntwo$ gauge theory, a detailed analysis of the temperature-induced
    deformation of moduli space has only been done for the case
    of an $SU(2)$ gauge group \cite{PaikYaf}, where it was argued
    that the discrete $\mathbb{Z}_2$ $R$-symmetry is spontaneously broken
    at low temperature.
    }
Hence, for finite $\Nc$, we do expect a distinct low temperature phase to exist
for some non-zero range of temperature.
However, it is quite plausible that the transition temperature scales as
$\Lambda/\Nc^\alpha$ for some $\alpha > 0$.
This would be consistent with our numerical results shown in figure~\ref{fig:free},
and is also consistent with the expected domain of validity of the low-energy
description near the degenerating points studied in ref.~\cite{Douglas}.
Further study of the low temperature behavior of this theory at
large but finite values of $\Nc$ would be interesting.

\section{Screening masses}
\label{sec:masses}

Using holography,
one may compute the thermal screening mass 
associated with a gauge invariant operator $\O$
by solving for the lowest eigenvalue
of the linear operator which describes fluctuations in the
corresponding dual field
\cite{Witten, Csaki, Koch, BMT}.
Let us remind the reader of the underlying logic.
The long-distance behavior of the
Euclidean two-point function of $\O$ determines
the screening mass.
If the Fourier transform of the correlator, viewed as a function
of the spatial wavevector, is analytic in a strip of width $\kappa$
above the real axis, then the coordinate space correlator will fall with
spatial separation $|\mathbf x {-} \mathbf x'|$ at least as fast as
$e^{-\kappa |\mathbf x {-} \mathbf x'|}$.
Hence, the screening mass $\kappa$ equals the distance from the real axis
to the nearest singularity.
If the Fourier transformed correlator is analytic
except for poles on the imaginary axis,
as will be the case in the large $\Nc$ theory under consideration,
then the associated screening mass equals the magnitude
of the pole closest to the origin.

There is a well-defined prescription for calculating 
Green's functions using gauge/gravity duality.
One must find solutions to the \emph{linearized} equations
of motion for the fluctuations of the supergravity field dual to $\O$.
We will use combinations of fluctuation fields which are
invariant under linearized diffeomorphisms.
After a 4D Fourier transform, fluctuation fields depend on
the radial coordinate $r$, the spatial wavevector $\mathbf k$,
and a discrete Matsubara frequency $\omega_n \equiv 2\pi n T$.
We will limit attention to zero frequency fluctuations,
since non-zero frequency fluctuations have shorter correlation lengths
than the corresponding static fluctuations.

The resulting linearized fluctuation equations of motion are
second order ordinary differential equations (ODEs) in $r$.
For some channels there is a single ODE involving just one fluctuation field;
for other channels there is a coupled set of ODEs involving multiple fields.
The ODE coefficients depend on the radial profiles of the 
scalar fields $\alpha$ and $\chi$ plus the 5D metric components 
describing the background geometry, as well as the fluctuation
wavevector $\mathbf k$ which appears in the dimensionless combination
$(\mathbf k L)^2$.

For the sake of discussion, consider 
a gauge invariant supergravity field $Z(r)$ that obeys a second order ODE.
A general solution may be expressed as a
linear combination of two independent solutions
having simple behavior near the horizon,
$
    Z(r) =
    c^\text{hor}_{\text I} \, Z_{\text{I}}^\text{hor}(r) +
    c^\text{hor}_{\text {II}} \, Z_{\text{II}}^\text{hor}(r)
$,
with coefficients $c_j^\text{hor}$ which
will depend on the wavevector $\bf k$.
The horizon is a regular singular point of the fluctuation ODE
and one solution, which we will choose to call $Z_{\text{II}}^\text{hor}$,
will have a logarithmic singularity at the horizon.
The physical solutions in which we are interested should be regular at
the horizon, and hence we must set $c_\text{II}^\text{hor} = 0$. 
The general solution may also be expressed as a linear combination
of two solutions having simple behavior near the boundary
($r \to \infty$),
$
    Z(r) =
    c^\text{bdy}_{\text I} \, Z_{\text{I}}^\text{bdy}(r) +
    c^\text{bdy}_{\text {II}} \, Z_{\text{II}}^\text{bdy}(r)
$.
We may choose independent solutions such that
$Z_{\text {I}}^\text{bdy}$ vanishes as $r \to \infty$
while $Z_{\text {II}}^\text{bdy} \to 1$ as $r \to \infty$.
Hence, the coefficient $c_\text{II}^\text{bdy}$ equals the
asymptotic value of the general solution $Z$.

The asymptotic value $c_\text{II}^\text{bdy}$
functions as the source for the
(Fourier-transformed) gauge theory operator $\O$, 
in accordance with the standard AdS/CFT interpretation.
Hence,
the Euclidean two-point function for $\O$ is the second
functional derivative of the regulated on-shell supergravity action
with respect to the value of $Z$ at the $r = \infty$ boundary.
The part of the on-shell action 
quadratic in fluctuations reduces to a boundary term of the form 
$\lim_{r \to \infty}\int d^3k\, \mathcal{F}(\mathbf{k}) Z'(r) Z(r)$ plus contact
terms which do not contain $Z'(r)$,
where the function
$\mathcal{F}(\mathbf{k})$ depends on details of the action \cite{KS}.
This leads to the result
$\vev{\O\O} \sim c_\text{I}^\text{bdy}/c_\text{II}^\text{bdy}$,
up to
terms analytic in $\mathbf{k}^2$ which can be removed by local counterterms.%
\footnote{%
The ratio $c_\text{I}^\text{bdy}/c_\text{II}^\text{bdy}$
is fixed by the required regularity of the solution at the horizon;
this ratio does not depend on the actual value of $c_\text{II}^\text{bdy}$.
}
Poles of the Green's function correspond to
zeros of $c_\text{II}^\text{bdy}$.
Consequently, to determine the screening mass one must find
the least negative value of $(\mathbf k L)^2$ for which the linearized
equation of motion for the fluctuation $Z$ has a solution satisfying
a Dirichlet condition at the boundary together with regularity at the horizon,
$c_\text{II}^\text{hor} = c_\text{II}^\text{bdy} = 0$.

We apply this procedure to evaluate the lightest screening masses
for the energy-momentum tensor
$T_{\mu\nu}$ of $\Ntwo^*$ SYM. This operator is dual to fluctuations of
the 5D metric in $\Neight$ gauged supergravity.  
We also obtain screening masses for the $\Nfour$ Lagrange density,
$\mathcal{L}_{\mathcal N=4} = \fourth \, \tr F^{\mu\nu}F_{\mu\nu} + \dotsb$,
dual to the 5D dilaton,
as well as the Pontryagin density,
$\tr E\cdot B = \fourth\, \tr F^{\mu\nu}\widetilde{F}_{\mu\nu}$,
which is dual to the 5D axion.

In each Euclidean correlator the operators are spatially separated
along the $x^3$ direction,
so field theory coordinates transverse to this longitudinal direction are
$x^1$, $x^2$, and $\tau$.
Distinct symmetry channels are classified
according to irreducible representations of the $O(2)$
group of rotations in the 1-2 plane,
together with discrete eigenvalues for
Euclidean time reversal (taking $\tau \to -\tau$),
which we will denote as $\mathcal R_\tau$,
and charge conjugation $\mathcal C$.
(The eigenvalues of $\mathcal R_\tau$
and $\mathcal C$ will be denoted by $R_\tau$ and $C$, respectively.)
Irreducible representations of $O(2)$ are labeled by a non-negative
integer angular momentum (or helicity) $\mathscr J$ plus, when $\mathscr J = 0$,
an eigenvalue for reflections in the 1-2 plane.
We will use $R_y$ to denote the eigenvalue of the reflection
$\mathcal R_y$ taking $x^2 \to -x^2$.
The resulting $\mathscr{J}^{C R_\tau}_{R_y}$ assignments
for various operators and supergravity fields
are conveniently tabulated in ref.~\cite{Bak}.

Using coordinates $x^\mu \equiv (\tau, x^1, x^2, x^3, z)$ with $L = 1$, 
the Euclidean black brane metric has the form
\begin{equation}
\label{background}
ds^2 = e^{2A(z)}\bigl[B(z)^2 \, d\tau^2 + d\mathbf{x}^2\bigr] + \frac{dz^2}{z^2}\,.
\end{equation}
We have switched to an inverted radial coordinate $z \equiv r_h/r$
which places the horizon at $z = 1$ and the boundary at $z = 0$.
At the horizon, $B(z)$ vanishes while $A(z)$ remains finite and non-zero.
Asymptotically, $A(z) \sim -\ln z$ and $B(z) \sim 1$ as $z \to 0$.
Appendix \ref{app:background} discusses the background geometry in more detail.

We add to this background 
a perturbation with arbitrary radial dependence
multiplying a plane wave in the $x^3$ direction with imaginary wavenumber $\kappa$.
In other words,
$g_{\mu\nu} = g^\text{cl}_{\mu\nu} + h_{\mu\nu}$, where $g^\text{cl}_{\mu\nu}$
is the background metric 
\eqref{background} and
\begin{equation}
h_{\mu\nu} \equiv h_{\mu\nu}(z) \, e^{-\kappa x^3} \,.
\end{equation}
We similarly expand
the two scalar fields present in the supergravity dual,
writing
$\alpha = \alpha_\text{cl} + \alphafluc$ and $\chi = \chi_\text{cl} + \chifluc$,
where the classical parts solve their respective Klein-Gordon equations
[with potential (\ref {eq:Vscalar})]
and the perturbations, indicated by tildes,
have the form 
\begin{equation}
\alphafluc \equiv \alphafluc(z) \, e^{-\kappa x^3}\,, \qquad
\chifluc \equiv \chifluc(z)\, e^{-\kappa x^3}\,.
\end{equation}
Under $O(2)$ rotations, the various fluctuation components
transform either as scalars ($\mathscr J = 0$),
vectors ($\mathscr J = 1$), or rank-2 tensors ($\mathscr J = 2$),
as shown in table~\ref{tab:polariz}.
The $O(2)$ symmetry guarantees that the equations for fluctuations
with differing helicities decouple.

An infinitesimal diffeomorphism, $\delta x^\mu = \xi^\mu$,
produces a metric perturbation 
$\delta h_{\mu\nu} = -\nabla_\mu\xi_\nu - \nabla_\nu\xi_\mu$,
along with variations of the
scalar fields given by
$\delta\alphafluc = -\xi^\mu \p_\mu\, \alpha_\text{cl}$ and
$\delta\chifluc = -\xi^\mu \p_\mu\, \chi_\text{cl}$.
The covariant derivative
$\nabla_\mu\xi_\nu \equiv \p_\mu\xi_\nu - \Gamma^\lambda_{\mu\nu}\,\xi_\lambda$
is taken with respect to the background metric.
Of particular relevance will be
residual diffeomorphisms which preserve
the functional form of the metric perturbation,
namely $\xi_\mu \equiv \xi_\mu(z) \, e^{-\kappa x^3}$.
It is helpful to note that there exist linear combinations
of metric perturbations (without radial derivatives)
which are invariant under these residual diffeomorphisms.
For example,
$h_{12}$ and $h_{0a}$ are residual diffeomorphism invariant perturbations
in the tensor and vector channels, respectively.
In the scalar channel,
$-h_{00} + \sum_{i=1}^3 \half(\Gamma^z_{00}/\Gamma^z_{ii}) \, h_+$ is invariant
under residual diffeomorphisms.
We will fix this residual diffeomorphism invariance by imposing the
``axial'' gauge condition,
\begin{equation}
\label{gauge}
    h_{\mu 4} = 0 \,,
\end{equation}
at all points in space.

\begin{table}[t]
\centerline{
\begin{tabular}{|c|c|l|}
\hline
$\mathscr J$ & mode & fields \\
\hline\hline
2 & tensor & $h_{ab} - \half h_+ \delta_{ab}$ \\
\hline
1 & vector & $h_{0a}$, $h_{a3}$, $h_{a4}$ \\
\hline
0 & scalar & $h_{00}$, $h_{03}$, $h_{04}$, $h_+$, $h_{33}$, $h_{34}$, $h_{44}$, $\alphafluc$, 
$\chifluc$ \\
\hline
\end{tabular}
}
\caption{Classifications of supergravity fields under $O(2)$ spatial rotations.
The index $a = 1,2$ labels transverse spatial directions,
3 is the longitudinal spatial direction,
and 4 the radial direction;
$h_+ \equiv h_{11} + h_{22}$.
\label{tab:polariz}}
\end{table}

The Einstein equations resulting from the action \eqref{action} are
\begin{equation}
\label{Einstein}
R_{\mu\nu} = 4\,(3\p_\mu \alpha \,\p_\nu \alpha + \p_\mu \chi \,\p_\nu \chi) 
+ \tfrac{4}{3}V(\alpha,\chi)\,g_{\mu\nu}\,,
\end{equation}
with the potential $V(\alpha,\chi)$ given by eq.~(\ref{eq:Vscalar}).
The scalar field equations are
\begin{equation}
\label{scalar}
\frac{1}{\sqrt{-g}}\,\p_\mu(\sqrt{-g} g^{\mu\nu}\p_\nu \alpha) = 
\tfrac{1}{6} \, V_\alpha \,,
\qquad
\frac{1}{\sqrt{-g}}\,\p_\mu(\sqrt{-g} g^{\mu\nu}\p_\nu \chi) = 
\tfrac{1}{2} \, V_\chi \,,
\end{equation}
where here and henceforth
we use subscripts to denote partial derivatives of the potential,
$V_\alpha \equiv \delta V/\delta \alpha$,
$V_\chi \equiv \delta V/\delta \chi$, etc.
Linearizing the Einstein equations about the background metric
$g_{\mu\nu}^\text{cl}$
produces the small fluctuation equation
\begin{align}
\label{linear_Einstein}
-\half\nabla_\mu\nabla_\nu h &-\half\nabla^2 h_{\mu\nu} 
+ \half\nabla^\lambda(\nabla_\mu h_{\nu\lambda} + \nabla_\nu h_{\mu\lambda})
\nonumber\\&=
\tfrac{4}{3}\bigl(
V h_{\mu\nu} 
+ V_\alpha \, g^\text{cl}_{\mu\nu} \alphafluc 
+ V_\chi \, g^\text{cl}_{\mu\nu} \chifluc 
\bigr)
+ 
12(\p_\mu\alpha_\text{cl} \, \p_\nu\alphafluc + \p_\mu\alphafluc \, \p_\nu\alpha_\text{cl})
\nonumber\\ & \hspace{2.16in}{}
+ 4(\p_\mu\chi_\text{cl} \, \p_\nu\chifluc + \p_\mu\chifluc \, \p_\nu\chi_\text{cl}) \,.
\end{align}
The linearized scalar fluctuation equations resulting from
eq.~\eqref{scalar} are
\begin{subequations}
\label{linear_scalar}
\begin{align}
\frac{1}{\sqrt{g_\text{cl}}}\,\p_\mu\left[
    \sqrt{g_\text{cl}} \left( \p^\mu\alphafluc -  h^{\mu\nu} \p_\nu \,\alpha_\text{cl} \right)
    \right]
+ \half\,\p_\mu h \, \p^\mu \alpha_\text{cl} &=
\tfrac{1}{6} \left(V_{\alpha\alpha}\,\alphafluc + V_{\alpha\chi}\,\chifluc \right),
\\
\frac{1}{\sqrt{g_\text{cl}}}\,\p_\mu\left[
    \sqrt{g_\text{cl}} \left( \p^\mu\chifluc -  h^{\mu\nu} \p_\nu \,\chi_\text{cl} \right)
    \right]
+ \half\,\p_\mu h \, \p^\mu \chi_\text{cl} &=
\half \left( V_{\chi\chi}\,\chifluc + V_{\chi\alpha}\,\alphafluc \right).
\end{align}
\end{subequations}
In these fluctuation equations, the background metric
$g^\text{cl}_{\mu\nu}$ is used to raise and lower indices,
compute covariant derivatives,
and define the trace $h \equiv g^{\mu\nu}_\text{cl}h_{\mu\nu}$.
Scalar potentials are evaluated on the classical solutions
$\alpha_\text{cl}$ and $\chi_\text{cl}$. 

Not counting the dilaton and axion, there are two equations for the scalars $\alpha$ and $\chi$ and fifteen equations for the 
components of the metric (which is a symmetric tensor).
In the axial gauge \eqref{gauge}, five of these equations become constraints 
and there are actually only five 
independent components of the metric, which split according to helicity into two for the tensor channel,
two in the vector channel, and one in the scalar channel --- which
couples to the scalar fields $\alpha$ and $\chi$.

\subsection*{Tensor channel}

\begin{figure}[t]
    \centerline{\includegraphics[scale=0.9]{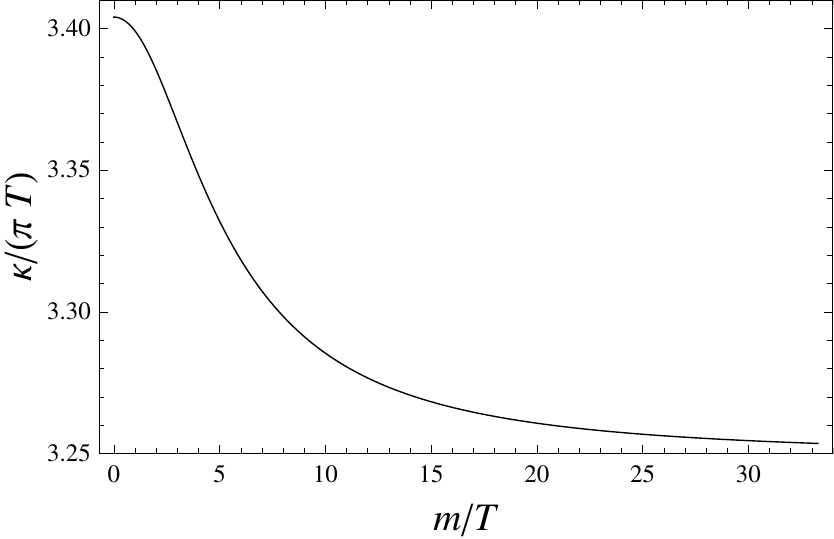}\quad\quad
    \includegraphics[scale=0.9]{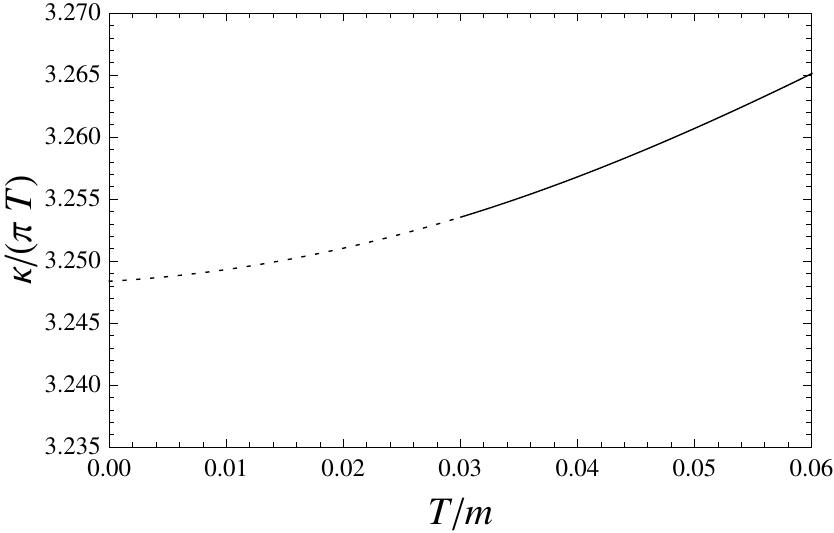}}
    \vspace*{-10pt}
    \caption{Screening mass (in units of $\pi T$) in the
             tensor channel,
	     $\mathscr{J}^{C R_\tau} = 2^{++}$,
	     as a function of $m/T$ (left)
	     and, for low temperatures, replotted as a function
	     of $T/m$ (right).
	     The dotted line is an
             extrapolation based on a quadratic fit to the data
	     (see footnote~\ref{foot}).
	     Fluctuations of the axion field, with
	     $\mathscr{J}^{C R_\tau} = 0^{+-}_-$,
	     and the dilaton, with
	     $\mathscr{J}^{C R_\tau} = 0^{++}_+$,
	     obey exactly the same linear equation as the $\mathscr J=2$
	     tensor mode,
	     and consequently
	     exhibit the same screening mass.}
    \label{fig:tensor}
\end{figure}

The tensor mode involves the symmetric traceless perturbation $h_{ab} - \half h_+ \delta_{ab}$.
Its two independent components are $h_- \equiv h_{11} - h_{22}$ and $h_{12}$,
which obey identical
second order ODEs.
The resulting equation may be written more compactly if we let
$b(z) \equiv \ln B(z)$,%
\footnote{%
This redefinition is not helpful for numerical calculations,
since $b(z)$ diverges at the horizon.
Nevertheless, we use it here to make the equations more concise.
}
and use redefined fields
$H_- \equiv e^{-2A} h_- $ and $H_{12}\equiv e^{-2A} h_{12} $.
The result is
\begin{equation}
H_-'' + (4A' + b' + z^{-1})\,H_-' + \kappa^2 \, z^{-2} {e^{-2A}}\,H_- = 0\,,
\end{equation}
(and likewise for $H_{12}$).
This is the same equation obeyed by the dilaton and axion --- 
they are `inert'
scalars which have constant background values and obey source-free Klein-Gordon equations 
\cite{DeWolfe}. Linearly-independent near-horizon solutions have the form 
\begin{equation}
H_-^\text{I}(z) \sim 1 + \dotsb, \qquad
H_-^\text{II}(z) \sim (1 + \dotsb) + H_-^\text{I}(z) \ln(1{-}z),
\end{equation}
where ellipses denote ascending power series in $z{-}1$.
As anticipated, the second solution
has a logarithmic singularity at the horizon, and must be discarded.
Since $e^{-2A}/z^2$ is regular at the horizon,
it follows that the appropriate boundary conditions at 
the horizon are $H_-(1) = c_0$ and $H_-'(1) = 0$,
with $c_0$ an arbitrary constant.
Linearly-independent near-boundary solutions have the form
\begin{equation}
H_-^\text{I}(z) \sim z^4\,(1 + \dotsb), \qquad
H_-^\text{II}(z) \sim (1 + \dotsb) + (\text{const.})\, H_-^\text{I}(z) \ln z\,,
\end{equation}
where these ellipses denote power series in $z$.
The coefficient of the non-normalizable solution $H_-^\text{II}$
must be set to zero, so that
$H_-$ vanishes as $z^4$ for small $z$. 
The numerical analysis of the resulting boundary value problem is
discussed in Appendix~\ref{app:tensor}.
The lowest screening mass obtained from this analysis is plotted
in figure~\ref{fig:tensor}
as a function of the ratio of the mass deformation to the temperature, $m/T$.
As $m/T \to 0$, we recover the $\Nfour$ value $\kappa \approx 3.4041 \, \pi T$.
Extrapolating the low temperature data shown on the right
in figure~\ref{fig:tensor}, we
find $\kappa/(\pi T) \to 3.248(5)$ as $T/m \to 0$.%
\footnote{%
\label{foot}
In performing the zero temperature extrapolation,
we assume that the screening mass 
admits a Taylor expansion of the form $\sum_{n \geq 0} p_n (T/m)^n$.
Truncating at quadratic order,
we used a least-squares fit to the data starting from the
smallest computed value of $T/m$. We looked for plateaus
in plots of the coefficients $p_n$ versus the number of fitted points, and
found a very stable fit using 125 points corresponding to
$0.0300 < T/m < 0.0315$. The leading
coefficient $p_0 \approx 3.2484$. 
Deviations were scattered randomly around zero 
with displacements on the order of $\pm 10^{-8}$.
Our estimate of the uncertainty in the last digit is based on
linear and cubic fits, which fit the data more poorly than
a quadratic fit and predict values
for $p_0$ which under- and over-shoot $3.2484$. 
}

\subsection*{Vector channel}

\begin{figure}[t]
    \centerline{\includegraphics[scale=0.9]{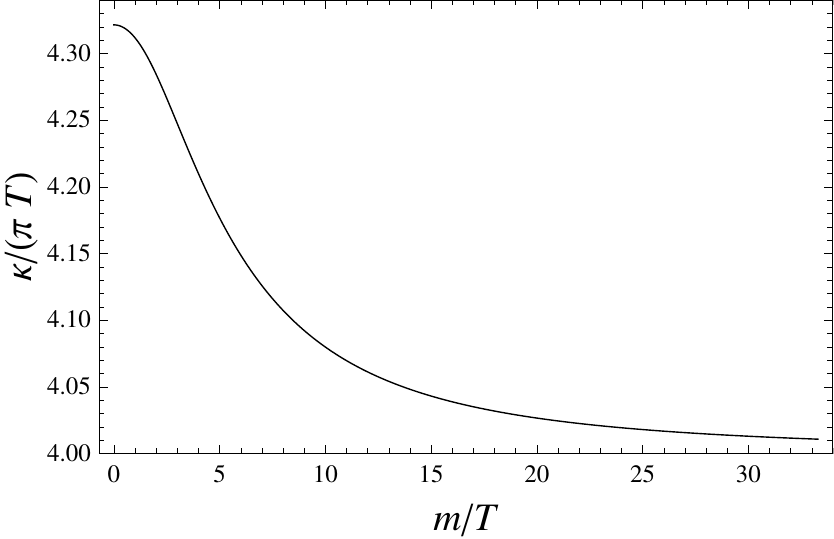}\qquad
    \includegraphics[scale=0.9]{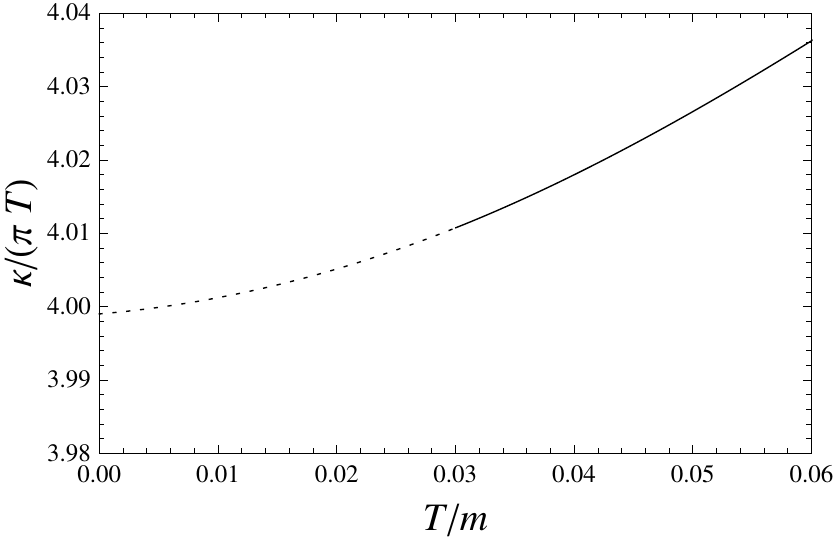}}
    \vspace*{-10pt}
    \caption{Screening mass (in units of $\pi T$) for  the
             vector channel, $\mathscr{J}^{C R_\tau} = 1^{+-}$,
             plotted as a function of $m/T$ (left) and, for low temperatures,
	     replotted as a function of $T/m$ (right).
             The dotted line is an
             extrapolation based on a quadratic fit to the data.}
    \label{fig:vector}
\end{figure}

In axial gauge the only independent vector fluctuation is $h_{0a}$.%
\footnote{%
The $h_{a4}$ component is zero from our gauge choice and the ($a4$) equations become a constraint that can be solved
as $h_{a3}(z) = C_{a4} e^{2A}$ for arbitrary 
constants $C_{a4}$.
Using the background equations of motion \eqref{ODEs}, it is straightforward
to show that this solution satisfies the second order ($a3$) equations.
The undetermined constant in $h_{a3}$ reflects the fact
that the axial gauge condition does not eliminate all diffeomorphism
freedom.
}
The $0a$ equations are fully decoupled.
After defining $H_{0a} \equiv e^{-2A}h_{0a} $, 
these equations take the form
\begin{equation}
H_{0a}'' + (4A' - b' + z^{-1})H_{0a}' + \kappa^2 z^{-2} {e^{-2A}}H_{0a} = 0\,.
\end{equation}
Near the horizon, linearly independent solutions are
\begin{equation}
H_{0a}^\text{I}(z) \sim (z{-}1)^2\,(1 + \dotsb), \qquad
H_{0a}^\text{II}(z) \sim (1 + \dotsb) + (\text{const.})\, H_{0a}^\text{I}(z) \ln (1{-}z).
\end{equation}
Both solutions are finite and once differentiable at $z = 1$.
Consequently, one cannot select a regular solution by specifying
a boundary value and first derivative precisely at the horizon.
However, one may use the regular near-horizon asymptotic solution
$H_{0a}^\text{I}$ to generate boundary conditions 
at some value of $z$ close to, but slightly less than 1.%
\footnote{%
Alternatively, one could remove the ambiguity by redefining 
$H_{0a}(z) \equiv (z{-}1)\Htilde_{0a}(z)$.
This shifts the roots of the indicial equation
down by one unit, from $\{2, 0\}$ to $\{1, -1\}$,
so that
the regular solution now behaves as $\Htilde_{0a}(z) \sim z-1$,
giving boundary conditions 
$\Htilde_{0a}(1) = 0$ and $\Htilde'_{0a}(1) = 1$, while the irregular solution diverges.
}
The near-boundary limit of the mode $H_{0a}$ is identical to that of $H_-$
since $b' \to 0 + O(z^3)$.
Once again, we demand that $H_{0a}$ vanish like $z^4$ as $z \to 0^+$.
Numerical implementation of this boundary value problem is identical
to that for the tensor mode.
Our results for the resulting vector
channel screening mass are plotted in figure~\ref{fig:vector}.
The limiting value for $m/T = 0$ is $\kappa \approx 4.3215 \, \pi T$.%
\footnote{%
This value is slightly smaller than the 
$\Nfour$ value quoted in ref.~\cite{Bak}
which was obtained in ref.~\cite{BMT}.
The discrepancy of $\approx 0.0002 \, \pi T$
presumably reflects less accurate numerical work in these references.
}
Extrapolating to zero temperature,
we estimate that $\kappa/(\pi T) \to 3.999(9)$ as $T/m \to 0$.

\subsection*{Scalar channel}

In axial gauge, the scalar channel couples the metric perturbations
$h_{00}$, $h_+ \equiv h_{11} + h_{22}$, and $h_{33}$ with
the scalar perturbations $\alphafluc$ and $\chifluc$.%
\footnote{%
The perturbation $h_{03}$ may be found by integrating the first order $(04)$
equation, leading to 
$h_{03}(z) = C_{03} B^2 e^{2A}$ for an arbitrary constant $C_{03}$. 
Using the background equations of motion \eqref{ODEs}, one may check
that this solution also satisfies the second order ($03$) equation.
}
The mixing of these fluctuations at nonzero $m/T$ is inevitable
--- the conformal
Ward identity given in eq.~\eqref{Ward} involves the thermal averages of
the operators $\O_2$ (dual to $\alphafluc$), $\O_3$ (dual to $\chifluc$),
and the trace $T^\mu_{~\mu}$ (dual to a
linear combination of $h_{00}$, $h_+$, and $h_{33}$).
It is convenient to make the following
field redefinitions. First, let $h_+ \equiv e^{2A}H_+$
and $h_{00} \equiv e^{2(A+b)}H_{00}$. Second, define
\begin{equation}
\label{gauge_invts}
Z \equiv H_{00} - \half(1+b'/A') \, H_+\,, \quad
a \equiv \alphafluc - \fourth(\alpha'_\text{cl}/A') \, H_+\,,  \quad
c \equiv \chifluc - \fourth(\chi'_\text{cl}/A') \, H_+ \,.
\end{equation}
These fields are invariant under infinitesimal diffeomorphisms preserving axial gauge.
In Appendix~\ref{app:scalar} we derive the coupled system of ODEs for
these gauge-invariant variables.
The result is
\begin{subequations}
\label{Zac}
\begin{align}
Z'' + p Z' + q Z &= f_a \, a + f_c \, c, \\
a'' + p_a a' + q_a a &= g_Z(Z' + b' Z) + g_c \, c, \\
c'' + p_c c' + q_c c &= h_Z(Z' + b' Z) + h_a \, a,
\end{align}
\end{subequations}
where
\begin{subequations}
\label{coefficient_fns}
\begin{align}
p &= 4A' + b' + z^{-1} - \tfrac 83 \, b' V/\mathcal Q
\,,\quad
f_a = \tfrac 83 \, {b'\left[(3A'{+}b')\,V_\alpha + 24\,\alpha'_\text{cl} V\right]}/\mathcal Q,
\\
q &= \kappa^2 z^{-2} {e^{-2A}} - \tfrac 83 \, {b'^2 V}/\mathcal Q
\,,\quad\quad
f_c = \tfrac 83 \, {b'\left[(3A'{+}b')\,V_\chi + 8\,\chi'_\text{cl} V\right]}/\mathcal Q,
\\[8pt]
p_a &= 4A' + b' + z^{-1}
\,,\qquad\qquad\quad\;\,
g_Z = \left( \tfrac 16 A' V_\alpha + \tfrac 43\,\alpha'_\text{cl} V \right)/\mathcal Q,
\\
p_c &= 4A' + b' + z^{-1}
\,,\qquad\qquad\quad\;\,
h_Z = \left( {\tfrac 12 A' V_\chi + \tfrac{4}{3}\chi'_\text{cl} V}\right)/ \mathcal Q ,
\\[8pt]
q_a &= \kappa^2 z^{-2} {e^{-2A}} - \tfrac 16 z^{-2} {V_{\alpha\alpha}} 
    - \tfrac 43 \left[{(6A'{+}b') \,\alpha'_\text{cl} V_\alpha}
    + {24\,\alpha'^2_\text{cl} \, V}\right]/\mathcal Q,
\\
q_c &= \kappa^2 z^{-2} {e^{-2A}}
    - \tfrac 12 z^{-2} {V_{\chi\chi}}
    - \tfrac 43 \left[{(6A'{+}b')\,\chi'_\text{cl} V_\chi}
    + {8\chi'^2_\text{cl} V} \right] / \mathcal Q ,
\\[8pt]
g_c &= \tfrac 16 z^{-2} {V_{\alpha\chi}}
    + \tfrac 43 \left[ {(3A'{+}b')\,\alpha'_\text{cl} V_\chi}
    + A' {\chi'_\text{cl} V_\alpha}
    + 8 \, {\alpha'_\text{cl}\chi'_\text{cl} V}\right]/\mathcal Q,
\\
h_a &= \tfrac 12 z^{-2} {V_{\alpha\chi}}
    + \tfrac 43 \left[ {(3A'{+}b')\chi'_\text{cl} V_\alpha}
    + {9 A'\alpha'_\text{cl} V_\chi}
    + {24\,\alpha'_\text{cl}\chi'_\text{cl} V}\right]/\mathcal Q\,,
\end{align}
\end{subequations}
with $\mathcal Q \equiv z^2 A'(3A'+b')$.%
\footnote{%
Equations (\ref{Zac})-(\ref{coefficient_fns})
are symmetric under the interchange $\sqrt{3}\,\alpha \leftrightarrow \chi$.
The factor of $\sqrt{3}$ reflects the normalization of the scalar kinetic
terms in the action (\ref{eq:Lmat}).
}

It is instructive to
first consider the massless limit, $m = 0$, corresponding to
the $\Nfour$ theory where we know that 
$\alpha_\text{cl} = \chi_\text{cl} = 0$.
This implies that $V = -\tfrac{3}{4}\hat{g}^2 = -3$, 
$V_\alpha = V_\chi = 0$, $V_{\alpha\alpha} = -6\hat{g}^2 = -24$, 
$V_{\chi\chi} = -\tfrac{3}{2}\hat{g}^2 = -6$, and $V_{\alpha\chi} = 0$.
All of the source 
terms on the right hand sides of eqs.~\eqref{Zac} vanish,
leaving fully decoupled ODEs.
The scalar fields $a$ and $c$ obey an equation of the form 
$\phi'' + \bigl(4A'+b'+1/z\bigr)\phi' + \bigl(\kappa^2 e^{-2A}/z^2 - M^2/z^2\bigr)\phi = 0$,
with $M^2 = -4$ and $-3$, respectively.
The leading behavior of linearly independent solutions is
\begin{subequations}
\label{ac_scaling}
\begin{align}
&\text{near horizon:}\, & 
a^\text{I}(z) &\sim 1 + \dotsb, \quad &
a^\text{II}(z) &\sim (1 + \dotsb) + a^\text{I}(z) \ln(1{-}z), 
\\
&& 
c^\text{I}(z) &\sim 1 + \dotsb, \quad &
c^\text{II}(z) &\sim (1 + \dotsb) + c^\text{I}(z) \ln(1{-}z),
\\[4pt]
&\text{near boundary:}\, &
a^\text{I}(z) &\sim z^2(1 + \dotsb), \quad & 
a^\text{II}(z) &\sim z^2(1 + \dotsb) + a^\text{I}(z) \ln z,
\\
&&
c^\text{I}(z) &\sim z^3(1 + \dotsb), \quad &
c^\text{II}(z) &\sim z(1 + \dotsb) + (\text{const.})\, c^\text{I}(z) \ln z.
\end{align}
\end{subequations}
The term $-M^2/z^2$ in the ODE is unimportant near $z = 1$,
so both scalars have the same 
near-horizon behavior. The metric perturbation field $Z$ obeys
$Z'' + \bigl(4A'+b'+z^{-1} + \frac{8b'}{2+(zA')^2}\bigr)Z' 
+ \bigl(\kappa^2 z^{-2} e^{-2A} + \frac{8b'^2}{2+(zA')^2}\bigr)Z = 0$.
The linearly independent solutions are
\begin{subequations}
\label{Z_scaling}
\begin{align}
&\text{near horizon:}\, &
Z^\text{I}(z) &\sim \frac {1 + \dotsb}{(z{-}1)^{2}}, &
Z^\text{II}(z) &\sim \frac {1 + \dotsb}{(z{-}1)^{2}} + Z^\text{I}(z) \ln(1{-}z), 
\\[4pt]
& \text{near boundary:}\, &
Z^\text{I}(z) &\sim z^4(1 + \dotsb), &
Z^\text{II}(z) &\sim (1 + \dotsb) + (\text{const.})\, Z^\text{I}(z) \ln z \,.
\end{align}
\end{subequations}

For numerical evaluations (at arbitrary $m/T$),
it is convenient to make field redefinitions
such that the unwanted solution (or its first derivative)
diverges at the horizon and boundary,
while the physical solution remains regular. 
Examining eqs.~\eqref{ac_scaling} and \eqref{Z_scaling}, one sees
that this can be accomplished by letting
\begin{equation}
\label{redef}
Z \equiv {z^3}{(z{-}1)^{-2}} \, \Ztilde \,, \qquad
a \equiv z^2\,\atilde \,, \qquad
c \equiv z^2\,\ctilde \,.
\end{equation}
It is also convenient to define the vector
\begin{equation}
\mathbf{W}(z) \equiv 
(\Ztilde',\, \atilde',\, \ctilde',\, \Ztilde,\, \atilde,\, \ctilde),
\end{equation}
in terms of which
eqs.~\eqref{Zac} become a system of six first order ODEs.
One can write this homogeneous 
system as a matrix equation of the form
\begin{equation}
\mathbf{W}' = \mathbf{L}(z;\kappa^2) \, \mathbf{W} \,,
\end{equation}
where $\mathbf{L}$ is
a $6 \times 6$ matrix involving the coefficients
\eqref{coefficient_fns} and field redefinitions \eqref{redef}.
For any choice of the vector $\mathbf W$ at the boundary,
the solution $\mathbf W(z)$ at an arbitrary point in the bulk is
linearly related,
$\mathbf{W}(z) = \mathbf{\Phi}(z;\kappa^2) \, \mathbf{W}(0)$,
where the transfer matrix $\mathbf\Phi$ depends on $\kappa^2$
and equals the identity at the boundary,
$\mathbf{\Phi}(0;\kappa^2) = \mathbf{1}$.
Expanding the solutions in power series near the horizon and boundary,
one can show that
\begin{subequations}
\label{BCs}
\begin{align}
\mathbf{W}(1) &= 
(-2\Ztilde_0,\, -2\atilde_0,\, -2\ctilde_0,\, \Ztilde_0,\, \atilde_0,\, \ctilde_0),
\\
\mathbf{W}(0) &= (\Ztilde_{0,0},\, 0,\, \ctilde_{0,0},\, 0,\, \atilde_{0,0},\, 0).
\end{align}
\end{subequations}
There are three undetermined coefficients at the horizon $(\Ztilde_0,\, \atilde_0,\, \ctilde_0)$ 
and three at the boundary $(\Ztilde_{0,0},\, \atilde_{0,0},\, \ctilde_{0,0})$.
In other words, at either endpoint there is a three-dimensional subspace
of solutions satisfying physical boundary conditions at that endpoint.
Our task is to find those values of $\kappa^2$ for which the
boundary-to-horizon transfer matrix maps some vector in the physical
subspace at the boundary into a vector lying in the physical subspace
at the horizon.
This may be accomplished by
testing linear dependence.
We choose basis vectors that span the physical subspaces,
\begin{subequations}
\begin{align}
&\text{horizon:}\, && \big\{
(-2, 0, 0, 1, 0, 0),\;
(0, -2, 0, 0, 1, 0),\;
(0, 0, -2, 0, 0, 1)\big\},
\\
&\text{boundary:}\, && \big\{
(1, 0, 0, 0, 0, 0),\;
(0, 0, 0, 0, 1, 0),\;
(0, 0, 1, 0, 0, 0)\big\}.
\end{align}
\end{subequations}
Each basis vector defines a set of initial conditions which may be
used to integrate the differential
equation from one endpoint, either $z = 0$ or 1,
to a matching point in the bulk, $z = z_*$.
This yields
a set of six vectors $\{\mathbf{W}(z_*)_{i=1,\dotsc,6}\}$,
where the first three are obtained by integrating from
the horizon and the last three are obtained by integrating from the boundary. 
For generic values of $\kappa^2$, these six vectors will be linearly
independent.
But if, for some value of $\kappa^2$, a solution exists
which satisfies the boundary conditions at both boundary and horizon,
then the solution may be expressed either as a linear combination of
$\{\mathbf{W}(z_*)_{i = 1,2,3}\}$, or as a linear combination of
$\{\mathbf{W}(z_*)_{i = 4,5,6}\}$.
In other words, for this value of $\kappa^2$,
the six vectors $\{\mathbf{W}(z_*)_{i=1,\dotsc,6}\}$
will not be linearly independent.
If we regard these vectors as columns of a $6 \times 6$ matrix,
\begin{equation}
\mathbf{D} \equiv
\begin{pmatrix}
\mathbf{W}(z_*)_1\,, &
\dotsc \,, &
\mathbf{W}(z_*)_6
\end{pmatrix},
\end{equation}
then $\det\mathbf{D}$ must vanish for values of $\kappa^2$
which correspond to screening masses.
Consequently, the task reduces to finding the roots of $\det \mathbf D$
as a function of $\kappa^2$.
Appendix~\ref{app:scalar} describes our
numerical procedures in somewhat more detail.

\begin{figure}[t]
    \centerline{\includegraphics[scale=0.9]{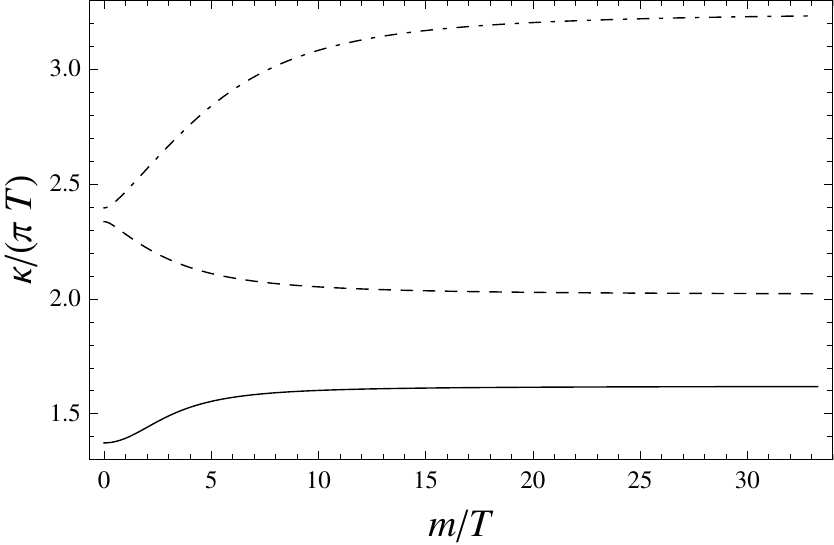}\qquad
    \includegraphics[scale=0.9]{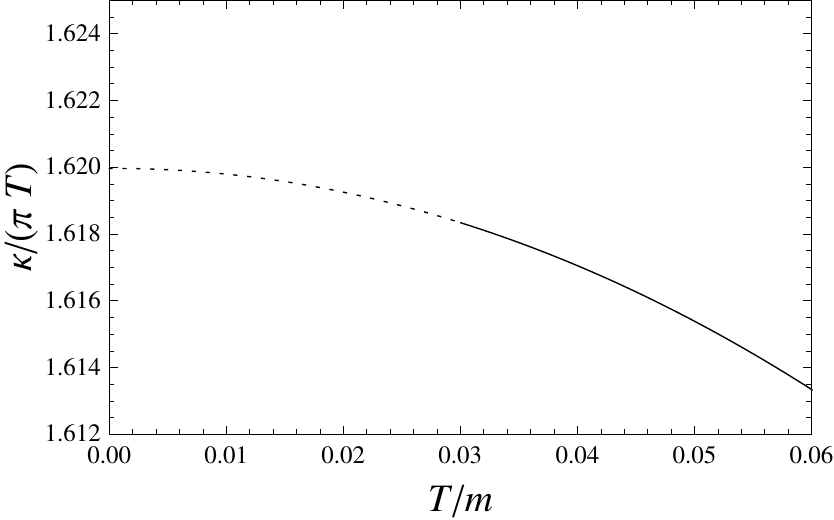}}
    \vspace*{-10pt}
    \caption{Screening masses (in units of $\pi T$) for the
             scalar channel, $\mathscr{J}^{C R_\tau}_{R_y} = 0^{++}_+$,
	     plotted as a function of $m/T$ (left).
	     For low temperatures, the lightest screening mass is
	     replotted (right) as a function of $T/m$.
             The dotted line is an extrapolation based on a
	     quadratic fit to the data.}
    \label{fig:scalar}
\end{figure}

Our results for the lowest screening masses are shown in figure~\ref{fig:scalar}. 
The operators dual to the scalar mode include the
energy density $T_{00}$ as well as $\O_2$ and $\O_3$. 
When $m/T=0$, the energy-momentum tensor and the scalar operators decouple.
At this point, the values shown
in figure~\ref{fig:scalar} correspond, in ascending order,
to the lowest screening mass of the $\O_2$, $T_{00}$,
and $\O_3$ operators in the undeformed $\Nfour$ theory.
The value of the second screening mass, $\kappa \approx 2.3361\, \pi T$,
associated with $T_{00}$ agrees with previous results \cite{Bak}.
For any nonzero value of $m/T$, these operators mix. Diagonalizing
the matrix of correlators will, at long distances,
yield decaying exponentials corresponding to the different
screening masses shown on the left in figure~\ref{fig:scalar}.
These curves may be regarded as the beginning
of a tower of possible screening masses.
The lightest screening mass is shown as a solid line, and its low
temperature behavior is replotted as a function of $T/m$
on the right side of figure~\ref{fig:scalar}.
Extrapolating to zero temperature,
we estimate that $\kappa/(\pi T) \to 1.620(2)$ as
$T/m \to 0$.

\section{Discussion}
\label{sec:compare}

Based on our results of screening masses
in $\Ntwo^*$ SYM plasma, we make the following
observations:
\begin{itemize} 
\item 
The smallest screening mass occurs in the $0^{++}_+$ channel for any value of $m/T$.
Fluctuations in the plasma with these quantum numbers are sourced by the energy density
$T_{00}$, as well as the scalar operators $\O_2$ and $\O_3$.
These operators couple to the time-time component of the
graviton and to quanta of the scalar fields $\alpha$ and $\chi$.
The associated screening mass is given by the solid 
curve in figure~\ref{fig:scalar} which rises monotonically (in $m/T$) from
$1.3731\, \pi T$ at high temperatures to $1.620\, \pi T$ at low
temperatures.
This screening mass should represent the mass gap, $\kappa_\text{gap}$, 
whose inverse characterizes the longest possible correlations in the plasma.%
\footnote{%
To settle the question of whether this really is the gap,
one should calculate the lightest screening masses in all possible 
symmetry channels.
We have not considered SYM operators coupling to the NSNS and RR
two-forms, since they have no simple 5D counterpart.
}
\item 
The smallest screening mass in an $R_\tau$-odd channel occurs in the $0^{+-}_-$ channel for any
value of $m/T$. It is shown in figure~\ref{fig:tensor} and is our best candidate for the Debye mass.
The associated supergravity mode is the 5D axion and its dual SYM operator is $\tr E\cdot B$.
\item 
Unlike the tensor and vector channels, the lightest screening mass in the scalar channel
increases as the temperature decreases. The mass gap in the weakly coupled theory exhibits a similar behavior in 
the range of temperatures $m\gg T\gg \Lambda$. The mass gap in units of the temperature depends quadratically on
the coupling constant  $m_{\rm gap}\sim g^2(T)\,T$,
and for temperatures $m\gg T$ the running of $g(T)$ 
is the same as for an asymptotically free theory. 
\item 
The screening mass extracted from the long-distance limit of the $T_{00}$ correlator jumps 
discontinuously when $m/T$ goes from zero to an infinitesimal positive number. 
This happens for the following reason. The Euclidean two point function can be
expressed as
$\vev{T_{00}(\mathbf{x}) T_{00}(0)} = \sum_n e^{-\kappa_n|\mathbf{x}|} |c_n|^2$, where 
$\{\kappa_n\}$ may be thought of as excitation energies for the eigenstates
$\{\ket{n}\}$ of the
Hamiltonian (or transfer matrix)
defined on $\mathbb{R}^2 \times S^1$, where the circumference of the compact spatial
direction is $\beta = 1/T$, and $c_n \equiv \vev{n|T_{00}|0}$. 
In the case of $\Nfour$ SYM, the lowest energy with a nonvanishing amplitude
to couple to $T^{00}$ is 
$\kappa_0 \approx 2.3361\, \pi T$ \cite{Bak}. However, when a mass deformation is turned on,
the $SO(6)_R$ symmetry of the massless theory
is explicitly broken to an $SU(2)_R$ subgroup.
Amplitudes $c_n$ which vanish at $m = 0$, due to the $SO(6)$ symmetry,
may become non-vanishing at $m \ne 0$,
when the $R$-symmetry is reduced.
From figure~\ref{fig:scalar} we learn that one of the corresponding 
energies is smaller than $\kappa_0$.
Thus, the behavior of the correlator in the $m/T \to 0$ and
$|\mathbf{x}| \to \infty$ limits depends on the order of limits.
Examining the long-distance behavior first,
for non-zero $m/T$,
yields a $T_{00}$ screening mass
which approaches $1.3731 \pi T$ as $m/T \to 0$.%
\footnote{%
We thank Andreas Karch for discussions about this point.
}
\end{itemize}

Let us comment on similarities and differences between
$\Nfour$ SYM, $\Ntwo^*$ SYM, and QCD plasmas:
\begin{itemize}
\item 
In QCD, the mass gap and the Debye mass are found in the $0^{++}_+$ and $0^{+-}_-$ channels, 
respectively.
This is also true for our candidate mass gap and Debye mass in $\Ntwo^*$ SYM.
And it is true for $\Nfour$ SYM with an important caveat:
one must limit the comparison of operators to the $R$-singlet sector.
It is known that the screening mass for an $R$-current in $\Nfour$ SYM 
is less than that for the energy density \cite{Amado}.
However, the $R$-current transforms
nontrivially under the $SO(6)_R$ of $\Nfour$ SYM.
QCD, of course, does not have such a symmetry,
which is why the authors of ref.~\cite{Bak} chose to restrict comparisons to
the $R$-singlet sector.
\item
In $\Nfour$ SYM, conformal invariance implies that screening masses are always proportional to the
temperature. In pure Yang-Mills theory at finite temperature, screening masses divided by the 
temperature are nearly constant for $1.5\Tc \leq T \lesssim 4\Tc$ \cite{Datta2,DG}.
Therefore, maximally
supersymmetric and non-supersymmetric non-Abelian plasmas are likely to be most similar in this
temperature window.
In $\Ntwo^*$ SYM, there are two regimes of
$m/T$ where screening masses scale linearly with temperature:
very high temperature, $m/T \ll 1$, where the mass deformation is
negligible,
and asymptotically low temperature, $m/T \to \infty$. 
These plateaus are clearly visible in, for example, figure~\ref{fig:scalar}.
One might expect the low temperature $\Ntwo^*$ plasma to be
most similar to QCD plasma at $T \approx 2\Tc$,
since the heavy $\Ntwo$ matter decouples in the low temperature regime
of $\Ntwo^*$ theory.

\begin{table}
\begin{center}

\begin{tabular}{|c|c|c|c|c|}
\hline
$\mathscr{J}^{C R_\tau}_{R_y}$ & $\Nf = 2$ QCD & $\Ntwo^*$ SYM & $\Nfour$ SYM 
\\ \hline\hline
$0^{++}_+$ & 1.25(2)  & 1.62 & 2.34 \\ \hline
$0^{+-}_-$ & 1.80(4)  & 3.25 & 3.40 \\ \hline
$1^{+-}$   & 2.88(12)\!\!\! & 4.01 & 4.32 \\ \hline
$2^{++}$   & 2.56(7)  & 3.25 & 3.40 \\ \hline
\end{tabular}
\end{center}
\caption{Screening masses in selected symmetry channels,
in units of $\pi T$,
in QCD ($\Nf = 2$, $T \approx 2\Tc$),
the large mass regime of $\Ntwo^*$ SYM ($m/T \approx 33.3)$,
and $\Nfour$ SYM.
\label{tab:masses}
}
\end{table}

In table \ref{tab:masses},
we show screening masses in various symmetry channels,
divided by $\pi T$,
in QCD [SU(3), $\Nf = 2$, $T \approx 2\Tc$] \cite{Hart},
$\Ntwo^*$ SYM in the large $m/T$ limit,%
\footnote
{More precisely, we show data from our largest mass value,
$m/T \approx 33.3$.}
and $\Nfour$ SYM.
As one sees from these results,
screening masses in QCD at $T \approx 2\Tc$ are not small
compared to $\pi T$.
This is a clear sign that QCD plasma in this regime is not weakly coupled.
Screening masses in $\Nfour$ SYM are roughly twice as large as
in $\Nf = 2$ QCD at $T \approx 2\Tc$.
Larger screening masses (or shorter screening lengths)
in $\Nfour$, relative to QCD,
is to be expected since there are many more fields
contributing to screening in $\Nfour$ theory.
As seen in table \ref{tab:masses}, screening masses in
the mass-deformed $\Ntwo^*$ theory, at large $m/T$,
are smaller than in $\Nfour$, and closer to the QCD values.
The decrease in screening masses, relative to $\Nfour$,
is relatively modest except for the $0^{++}_+$ scalar channel,
where the change is substantial.

We remind the reader that the $0^{+-}_-$ and $2^{++}$ screening
masses are necessarily identical in both
$\Nfour$ and $\Ntwo^*$ SYM.
This reflects the fact that in both these theories
the axion obeys the same equation of motion as the transverse components
of the graviton. A holographic theory in which these two supergravity
modes obey distinct equations will be needed to more closely model QCD.

\item

It is also interesting to compare ratios of screening masses in
different symmetry channels, instead of their absolute values.
Figure \ref{fig:ratios} plots screening mass ratios,
relative to the mass gap (or screening mass in the $0^{++}_+$
channel) as a function of $m/T$ in $\Ntwo^*$ SYM.
Table \ref{tab:ratios} shows the ratios of screening masses
relative to the mass gap for
QCD [$\Nf = 2$, $T \approx 2\Tc$] \cite{Hart},
$\Ntwo^*$ SYM in the large $m/T$ limit,
and $\Nfour$ SYM.
The ratio of the $0^{+-}_-$ and $0^{++}_+$ screening masses
is virtually identical in QCD and $\Nfour$ SYM.
This near-perfect agreement is surely fortuitous,
and in this particular case
the $\Ntwo^*$ ratio deviates much farther from the QCD value.
But, as shown in table \ref{tab:ratios},
in the other symmetry channels for which we have results,
the $\Ntwo^*$ screening mass ratios are much closer
to the QCD values than the $\Nfour$ ratios.

\begin{figure}[t]
    \centerline{\includegraphics[scale=0.8]{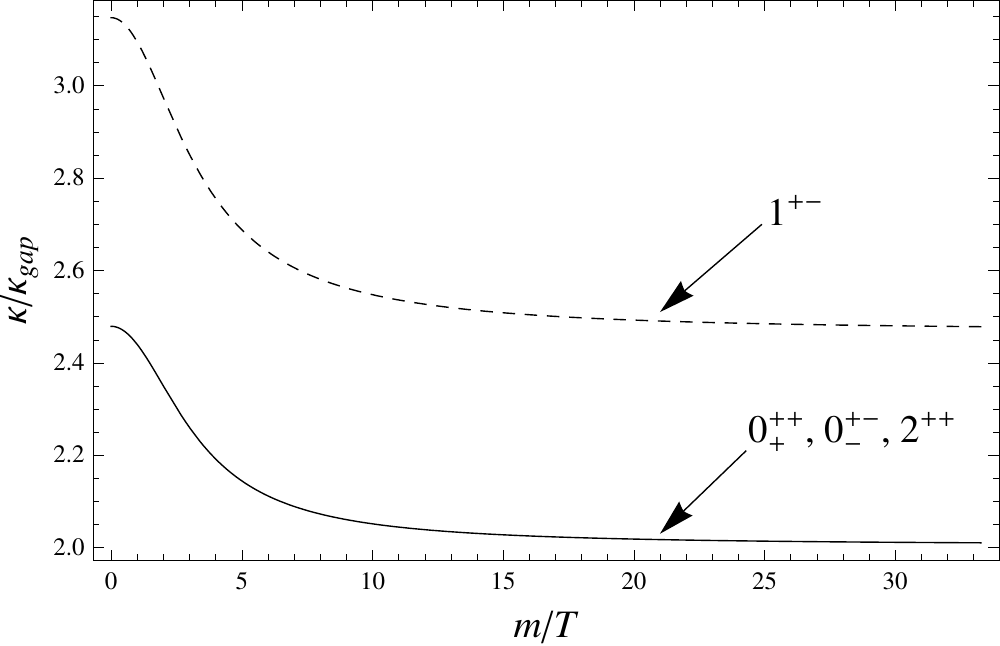}}
    \vspace*{-10pt}
    \caption{Ratios of $\Ntwo^*$ screening masses (in indicated channels)
    relative to the mass gap (or screening mass in the $0^{++}_+$ channel),
    as a function of $m/T$.}
    \label{fig:ratios}
\end{figure}

\begin{table}
\begin{center}

\begin{tabular}{|c|c|c|c|c|}
\hline
$\mathscr{J}^{C R_\tau}_{R_y}$ & $\Nf = 2$ QCD & $\Ntwo^*$ SYM & $\Nfour$ SYM 
\\ \hline\hline
$0^{+-}_-$ & 1.45(2)  & 2.01 & 1.46 \\ \hline
$1^{+-}$   & 2.31(10) & 2.48 & 1.85 \\ \hline
$2^{++}$   & 2.05(6)  & 2.01 & 1.46 \\ \hline
\end{tabular}
\end{center}
\caption{Ratios of screening masses in the indicated symmetry
channel to the mass gap (or screening mass in the $0^{++}_+$ channel),
in QCD ($\Nf = 2$, $T \approx 2\Tc$),
the large mass regime of $\Ntwo^*$ SYM ($m/T \approx 33.3)$,
and $\Nfour$ SYM.}
\label{tab:ratios}
\end{table}

Overall, considering both the absolute values and the ratios of
screening masses in various symmetry channels,
it seems fair to regard $\Ntwo^*$ SYM as a better model
of a QCD plasma than $\Nfour$ SYM.

\item
In pure Yang-Mills theory,
screening masses divided by the temperature drop precipitously 
(at least in certain $\mathscr{J} = 0$ channels)
when the temperature falls below $1.5\Tc$ and
approaches $\Tc$ \cite{DG}.
Neither $\Nfour$ nor $\Ntwo^*$ SYM have thermal
phase transitions at a non-zero temperature.
(For the former this is certain, and for the latter we find no evidence.)
Thus, it is not sensible to ask how well either supersymmetric
theory models QCD near $\Tc$.
However, as in QCD, screening masses of $\Ntwo^*$ SYM
change substantially over a relatively small temperature range.
From figures \ref{fig:tensor}--\ref{fig:ratios},
it is clear that the high temperature plateau
is limited to $m/T \lesssim 1$ and the low temperature plateau is
roughly $m/T \gtrsim 10$
---
significant variation is confined to the band $m/T \approx \mathrm{few}$.
The point $m/T \approx 4.83$, where the trace
anomaly deviates maximally from zero (c.f. figure~\ref{fig:trace}),
lies in the middle of this band.

\end{itemize}

\acknowledgments

We are grateful to Andreas Karch for numerous helpful discussions.
We have also benefited from conversations with Alex Buchel
and Kristan Jensen. 
S.P. thanks Brian Smigielski for assistance with the {\it Condor}
computing software
implemented at the University of Washington Physics Department.
This work was supported in part by the U.S. Department of Energy
under grant DE-FG02-96ER40956.

\appendix

\section{Properties of the gravity dual}
\label{app:background}

In this and the following appendices, we present details of our
analytical and numerical calculations at a level which would allow someone,
with some effort, to reproduce the results presented above.
Further details, including explicit expressions for the coefficients
of series expansions and more explicit derivations, may be obtained
online \cite{online}.

This appendix summarizes the 5D $\Neight$ supergravity solution dual to finite temperature 
large-$\Nc$ $\Ntwo^*$ SYM. Most of the discussion is taken directly from the original works
\cite{BLthermo, Bhydro, BDKL}. We collect the results in one place for two reasons: to explain,
in detail, how our calculations were performed, and to benefit readers wishing to do further
studies.
We used {\it Mathematica 7.0} for both analytical and numerical computations.

\subsection*{Equations of motion}

We take the following ansatz for the 5D black brane metric,
\begin{equation}
\label{metric}
ds^2 = e^{2A(r)} \bigl(-B(r)^2 dt^2 + d\mathbf{x}^2\bigr) + dr^2.
\end{equation}
Without loss of generality we have chosen coordinates such that $g_{rr} = 1$ everywhere. The
metric is invariant under 3d spatial rotations, as well as time and spatial translations. The 
spacetime is defined for values of the radial coordinate $r_h \leq r < \infty$. The lower end is
an event horizon with the property that $B(r_h) = 0$, and $A(r_h) = A_h$, a finite number. Infinity
is a boundary where the metric approaches that of $AdS_5$. This requires 
$\lim_{r\to\infty}B(r) = 1$ and $A(r)$ to asymptotically approach $r/L$.%
\footnote{%
The asymptotic behavior of $A(r)$ could also be $r/L + A_\infty$, where $A_\infty$ is an arbitrary
constant. However, $A_\infty$ can be set to zero by rescaling the time coordinate.
}
The scalar fields are chosen to be functions of $r$ only. They should be
finite at the horizon: $\alpha(r_h) = \alpha_h$ and $\chi(r_h) = \chi_h$, and vanish at infinity:
$\lim_{r\to\infty} \alpha(r) = \lim_{r\to\infty} \chi(r) = 0$. We can shift $r$ and compensate the change in the asymptotic form of the metric by rescaling the spacetime coordinates. Let us then choose $r_h = 0$ without loss of generality.
The radial coordinate $r$ is not
well-suited to numerical calculations since the boundary is at infinity. Let us switch to a new,
dimensionless radial coordinate $z \equiv e^{-r/L}$.%
\footnote{%
Some derivations are more transparent using the $r$ coordinate, whereas technical calculations are
better handled using the $z$ coordinate. We will freely switch back and forth
between $r$ and $z$ in the following.
}
This places the horizon at $z = 1$ and the boundary at $z = 0$. In this new coordinate system, 
$A(z) \sim -\ln z$ as $z \to 0$. Since the logarithmic divergence is troublesome for numerical 
work it is helpful to extract this asymptotic behavior by defining
\begin{equation}
A(z) \equiv -\ln z + \Atilde(z).
\end{equation}
Note that this field redefinition does not significantly alter
the behavior of the warp factor $A$ near the horizon. In particular, $\Atilde(1) = A_h$.
The metric now reads
\begin{equation}
\label{metricz}
ds^2 = \frac{e^{2\Atilde(z)}}{z^2}\bigl( -B(z)^2 dt^2 + d\mathbf{x}^2 \bigr)+L^2 \, \frac{dz^2}{z^2}.
\end{equation}
The Einstein equations derived from the action \eqref{action} and metric
\eqref{metricz} are
\begin{subequations}
\label{ODEs}
\begin{align}
& B' = C z^3 e^{-4\Atilde}, \label{constraintB} \\
& \bigl(\Atilde' - 1/z\bigr)^2 + \half\bigl(\Atilde' - 1/z\bigr)B'/B -\alpha'^2 -\third \chi'^2 
+ \frac{L^2 V}{3z^2} = 0, \label{constraintA} \\
& \alpha'' + \bigl(4\Atilde' + B'/B - 3/z\bigr)\alpha' - \frac{L^2 V_\alpha}{6z^2} = 0, \\
& \chi'' + \bigl(4\Atilde' + B'/B - 3/z\bigr)\chi' - \frac{L^2 V_\chi}{2z^2} = 0,
\end{align}
\end{subequations}
where a prime denotes $d/dz$ and $C$ is a constant that will be determined later. 
The system \eqref{ODEs} is sixth order.%
\footnote{%
By making the clever choice of radial coordinate $t(z) \equiv 1-B(z)$, the authors of 
ref.~\cite{BDKL} reduce the system to a set of three second order ODEs which does not explicitly 
contain $B$ or its derivative. We decided not to use the $t$ coordinate because it leads to 
near-horizon and near-boundary expansions with fractional powers of $t$;
we find the series 
solutions simpler to understand in the $z$ coordinate.
}
Note that factors of $L^2$ appearing in the system and in the gauged supergravity coupling present 
in the scalar potential cancel. Therefore, it is natural to set $L = 1$ and $\hat{g}^2 = 4$.%
\footnote{%
This is different from the convention used in ref.~\cite{BDKL} where $L = 2$ and $\hat{g}^2 = 1$.
}

\subsection*{Near-horizon solution: temperature and entropy}
\label{app:temptropy}

To have a regular black brane horizon, near $z = 1$, we try a power series solution of the form
\begin{subequations}
\label{nearhorsol}
\begin{align}
B(z) &= -B_h(z{-}1)\Bigl[1 + \sum_{n \geq 1} B_n (z{-}1)^n\Bigr], \label{Bhor} \\
\Atilde(z) &= A_h + \sum_{n \geq 1} \Atilde_n (z{-}1)^n, \label{Ahor}
\\
\alpha(z) &= \alpha_h + \sum_{n \geq 1} \alpha_n (z{-}1)^n, \\
\chi(z) &= \chi_h + \sum_{n \geq 1} \chi_n (z{-}1)^n.
\end{align}
\end{subequations}
Substituting these into system \eqref{ODEs} (first taking the logarithm of eq.~\eqref{constraintB} 
then differentiating with respect to $z$),
we can solve the equations order-by-order in $z-1$. We solve up to sixth order.
The asymptotic values $A_h$, $B_h$, $\alpha_h$, and $\chi_h$ are undetermined,
and will be referred to as ``horizon data.''
All higher coefficients $\{B_n, \Atilde_n, \alpha_n, \chi_n\}$
are fully determined by the horizon data.
At the horizon, the fields and their slopes are finite, as desired,
\begin{subequations}
\begin{align}
B(1) &= 0, & B'(1) &= -B_h, & \alpha(1) &= \alpha_h, & \alpha'(1) &= 0, \\
\Atilde(1) &= A_h, & \Atilde'(1) &= 1, & \chi(1) &= \chi_h, & \chi'(1) &= 0.
\end{align}
\end{subequations}

The Hawking temperature of the black brane may be extracted from the metric near the horizon using the usual analytical continuation to Euclidean signature with a compact time direction, and demanding that the solution is regular at the horizon,  where the time circle collapses to zero size. This fixes the period of the temporal circle to be
\begin{equation}
\label{dict:temperature}
\beta=\frac{2\pi L}{ e^{A_h}B_h }.
\end{equation}
In the AdS/CFT correspondence, we identify the black brane temperature $T = 1/\beta$ with the 
temperature in the quantum field theory.

The Bekenstein-Hawking entropy of the black brane is proportional to the volume of its 
three-dimensional horizon divided by Newton's constant in five dimensions, 
$S = V_\text{horizon}/(4G_5)$. The horizon volume is $V_\text{horizon} = \int_{\mathbb{R}^3} d^3x\,
\sqrt{\det g^\text{ind}_{ab}}$, where $g^\text{ind}_{ab} = e^{2A_h}\delta_{ab}$ is the induced
metric on the hypersurface $z = 1$ and fixed $\tau$. Therefore, $V_\text{horizon} = \vol\, e^{3A_h}$, 
where $\vol$ is the (formally infinite) volume of $\mathbf{x}$-space. The entropy density is
\begin{equation}
\label{dict:entropy}
S/\vol = \frac{e^{3A_h}}{4G_5} = \frac{e^{3A_h}\Nc^2}{2\pi L^3} = \frac{4\pi^2 \Nc^2 T^3}{B_h^3}.
\end{equation}
Writing this as in eq.~\eqref{entropy}, we identify $\sigma \equiv (2/B_h)^3$. Positivity of
entropy implies that $B_h > 0$.
We can fix the unknown constant $C$ in eq.~\eqref{constraintB} using the near-horizon solution. 
Plugging eqs.~\eqref{Bhor} and \eqref{Ahor} into eq.~\eqref{constraintB}, then evaluating at
$z = 1$, we obtain $C = -e^{4A_h}B_h$.

\subsection*{Near-boundary solution: gauge-gravity dictionary}
\label{app:dict}

Near $z = 0$, solutions may be expanded in an asymptotic series solution of the form
\begin{subequations}
\label{nearbdysol}
\begin{align}
B(z) &= 1 + \sum_{n= 0}^\infty \sum_{k = 0}^{n} B_{2n+4,k} \, z^{2n+4} (\ln z)^k, \label{Bbdy}
\\
\Atilde(z) &= \sum_{n= 0}^\infty \sum_{k = 0}^{n+1} \Atilde_{2n+2,k} \, z^{2n+2} (\ln z)^k, \label{Abdy}
\\
\alpha(z) &= 
\sum_{n= 0}^\infty \sum_{k = 0}^{n+1} \alpha_{2n,k}\, z^{2n+2} (\ln z)^k,
\\
\chi(z) &= 
\sum_{n= 0}^\infty \sum_{k = 0}^n \chi_{2n,k}\, z^{2n+1} (\ln z)^k.
\end{align}
\end{subequations}
The leading (small $z$) behavior of the fields $B$ and $\Atilde$ are dictated by the requirement 
that, at the boundary, the metric approaches that of $AdS_5$.
Moreover, the two scalar fields must
vanish at the boundary. The rate at which they vanish ({i.e.}, the leading exponent) is 
determined by linearizing the scalar equations of motion at $z = 0$ and solving the characteristic 
equation for an ansatz of the form $z^p$.%
\footnote{%
The scalar potential is, to quadratic order 
in the small fields, $-L^2 V \approx 3 + 12\alpha^2 + 3\chi^2$. Therefore, the linearized scalar
equations are $\alpha'' - (3/z)\alpha' + (4/z^2)\alpha = 0$ and $\chi'' - (3/z)\chi' + (3/z^2)\chi
= 0$. The first equation has $p = 2$ as a double root. Therefore, a pair of linearly independent
solutions is $z^2$ and $z^2\ln z$. The second equation has $p = \{3,1\}$, so a pair of 
linearly independent solutions is $z^3$ and $z + c z^3\ln z$, with $c$ a nonzero
coefficient determined by the equations of motion.
}
The presence of $\ln z$ in the leading series solution comes from the fact that $z = 0$ is a
regular singular point and the roots of the characteristic equation
differ by an integer. Higher powers of $\ln z$
arise from nonlinearities in the system \eqref{ODEs}.
The unknown coefficients are found by plugging the ansatz into the system
\eqref{ODEs} and solving order-by-order in $z^a(\ln z)^b$.
We carried this through up to order $a = 8$.
Some coefficients vanish, $A_{2,1}=B_{6,1}=0$,
but generically all terms in the series \eqref{nearbdysol} are non-zero.
Observe that
eq.~\eqref{constraintB} determines the $B_{n,k}$ in terms of the $\Atilde_{n,k}$, and 
eq.~\eqref{constraintA} determines the $\Atilde_{n,k}$ in terms of the $\alpha_{n,k}$ and 
$\chi_{n,k}$. The first order equations \eqref{constraintB} and \eqref{constraintA} 
should be thought of as constraints.
The two scalar equations determine all higher 
$\alpha_{n,k}$ and $\chi_{n,k}$ in terms of the first couple of coefficients. 
All higher coefficients $\{B_{n,k}, \Atilde_{n,k}, \alpha_{n,k}, \chi_{n,k}\}$ 
are fully determined by the ``boundary data":
$B_{4,0}$, $\alpha_{0,1}$, $\alpha_{0,0}$, $\chi_{0,0}$, and $\chi_{2,0}$. 
At the boundary, the fields and their slopes are
\begin{equation}
\begin{split}
B(0) = 1, \quad B'(0) = 0, &\qquad \alpha(0) = 0, \quad \alpha'(0) = 0, \\
\Atilde(0) = 0, \quad \Atilde'(0) = 0, &\qquad \chi(0) = 0, \quad \chi'(0) = \chi_{0,0}.
\end{split}
\end{equation}
We can fix the constant $C$ in eq.~\eqref{constraintB} using the near-boundary solution.
In fact, we may evaluate this equation at any value of $z$ and the constant $C$ must be the 
same --- doing this at the horizon and boundary provides a powerful constraint.
Plugging eqs.~\eqref{Bbdy} and \eqref{Abdy} into eq.~\eqref{constraintB}, dividing by $z^3$, 
then evaluating at $z = 0$, yields $C = 4B_{4,0}$. We now have a nontrivial equation
that relates boundary and horizon data,
\begin{equation}
\label{nontrivial}
4B_{4,0} = -e^{4A_h} B_h.
\end{equation}
To summarize, the near-boundary solutions are
\begin{subequations}
\begin{align}
B(z) &= 1 + B_{4,0}\, z^4 + O(z^6), \\
\Atilde(z) &= \Atilde_{2,0}\, z^2 + O(z^4\ln^2 z), \\
\alpha(z) &= z^2\bigl[\alpha_{0,1}\ln z + \alpha_{0,0} + O(z^2\ln^2 z)\bigr], \\
\chi(z) &= z\bigl[\chi_{0,0} + z^2(\tfrac 43 \chi_{0,0}^3\ln z + \chi_{2,0}) + O(z^4\ln^2 z)\bigr].
\end{align}
\end{subequations}
Generally, in the AdS/CFT correspondence, the leading coefficient in a near-boundary expansion
corresponds to the source for the dual operator and the subleading coefficient corresponds to the
expectation value of that operator. The operators dual to $\alpha$ and $\chi$ are 
$\O_2$ and $\O_3$, respectively. Hence $\alpha_{0,1}$ will be proportional to $m^2$ and 
$\alpha_{0,0}$ is related to the thermal expectation $\vev{\O_2}$. Likewise, $\chi_{0,0}$ will
be proportional to $m$ and $\chi_{2,0}$ related to $\vev{\O_3}$.

The precise correspondence relies on a matching calculation at zero temperature. At $T = 0$
the background metric must be Poincar\'e-invariant. This forces $B = 1$. The remaining fields
$A$, $\alpha$, and $\chi$ obey a system of 1st order ODEs referred to as the supersymmetric
flow equations \cite{PW}, whose solution is
\begin{subequations}
\label{susyflow}
\begin{align}
e^A &= k\, {e^{2\alpha}}/{\sinh 2\chi}, \label{susyflow1} \\
e^{6\alpha} &= \cosh 2\chi + (\gamma + \ln \tanh \chi)\sinh^2(2\chi), \label{susyflow2}
\end{align}
\end{subequations}
where $k$ and $\gamma$ are constants of integration. Plugging in the asymptotic formulas for the 
fields fixes the undetermined coefficients. At leading order,
\begin{equation}
\label{leading}
\alpha_{0,1} = \tfrac{2}{3}\chi_{0,0}^2, \qquad \chi_{0,0} = k/2.
\end{equation}
At the first subleading order one finds
\begin{equation}
\label{subleading}
\alpha_{0,0} = \third\chi_{0,0}^2\,(1 + 2\gamma + 2\ln\chi_{0,0}), \qquad
\chi_{2,0} = 2\alpha_{0,0}\,\chi_{0,0} - \third\chi_{0,0}^3.
\end{equation}
The constant $\gamma$ parameterizes a family of distinct solutions. Solutions with 
$\gamma \leq 0$ correspond to points on the Coulomb branch of $\Ntwo^*$ SYM (where the scalars 
$\phi_1$ and $\phi_2$ are massive and $\phi_3$ gets an expectation value). Solutions with $\gamma > 0$
are unphysical. The expressions given in eq.~\eqref{leading} for the leading
series coefficients should not change at finite temperature since they are bare couplings
in a Lagrange density. As such, the constant $k$ must equal $mL$ times a pure number.
Eq.~(\ref{leading}) must also hold for the black brane solution. However, eq.~\eqref{subleading} 
holds only at zero temperature. For the case $\gamma = 0$, a probe D3-brane computation shows that 
the supergravity solution may be understood as a Coulomb branch 
vacuum in which the background D3-branes form a linear enhan\c{c}on singularity. A computation of
the size of the enhan\c{c}on shows that $k = mL$. Thus,
\begin{equation}
\label{dict:mass}
\chi_{0,0} = \frac{mL}{2}, \qquad \alpha_{0,1} = \frac{(mL)^2}{6}.
\end{equation}

\begin{figure}[t]
    \centerline{\includegraphics[scale=0.9]{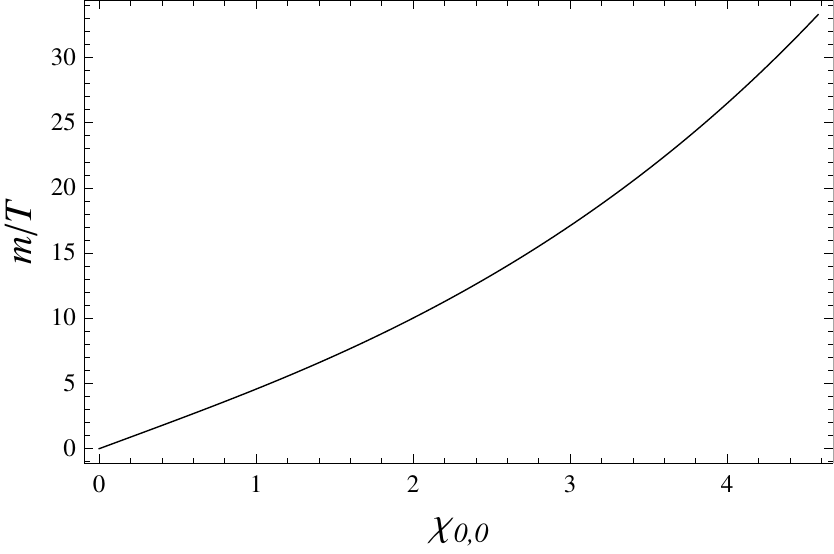}}
    \vspace*{-10pt}
    \caption{$m/T$ as a function of $\chi_{0,0}$.}
    \label{fig:mbeta}
\end{figure}
We choose to regard $\chi_{0,0}$ as an independent variable on which the other supergravity
coefficients depend. In other words, the boundary and horizon data are all
functions of $\chi_{0,0}$:
\begin{equation}
\label{data}
B_{4,0}(\chi_{0,0}), \quad
\alpha_{0,0}(\chi_{0,0}), \quad
\chi_{2,0}(\chi_{0,0}), \quad
A_h(\chi_{0,0}), \quad
\alpha_h(\chi_{0,0}), \quad
\chi_h(\chi_{0,0}).
\end{equation}
There are six independent data --- $\alpha_{0,1}$ is fixed by eq.~\eqref{dict:mass} and $B_h$ is
fixed by eq.~\eqref{nontrivial}.

Solving for $L$ in eq.~\eqref{dict:temperature}, the map from 
$\chi_{0,0}$ in eq.~\eqref{dict:mass} to $m/T$  is
\begin{equation}
\label{mbeta}
m/T = \frac{4\pi\,\chi_{0,0}}{e^{A_h}B_h}.
\end{equation}
A plot of this map is shown in figure~\ref{fig:mbeta}. The mapping is nonlinear and monotonic, with 
larger values of $\chi_{0,0}$ corresponding to larger values of $m/T$. Since $A_h(0)$ and $B_h(0)$ 
are nonzero (see Appendix~\ref{app:numerics}), $m/T = 0$ when $\chi_{0,0} = 0$.

\subsection*{Holographic renormalization, thermodynamics and expectation values}
\label{app:holorg}

The Euclidean supergravity action is regulated and renormalized
following the 
approach of ref.~\cite{Bhydro}.
Proper renormalization is needed to define 
thermal expectation values correctly.
We start by cutting off the boundary at the hypersurface given by the equation $r = r_0$. 
To the bulk Euclidean action $S_\text{bulk}$ given by (minus) 
eq.~\eqref{action}, we add the Gibbons-Hawking boundary term
\begin{equation}
S_\text{bdy} = -\frac{1}{8\pi G_5}\int_{r = r_0}d^4x\, \sqrt{h} K,
\end{equation}
where $h$ is the determinant of the induced metric $h_{\mu\nu}$ on the hypersurface, and $K$ is the
trace of the extrinsic curvature. The sum of these actions will diverge in the limit that 
$r_0 \to \infty$. To cure this one must add counterterms at the boundary. When the boundary is flat the counterterm action has the general form
\begin{equation}
\label{counterterm}
S_\text{ctr} =
\frac{\beta V}{4\pi G_5} \,
e^{4A}B\Bigl(
c_1 + c_3\, \alpha + c_4\,\chi^2 + c_5\,\alpha^2 + c_6\,\alpha\chi^2 + c_8\,\frac{\alpha^2}{\ln x} 
+ c_{10}\,\chi^4\ln x + c_{11}\,\chi^4\Bigr)\biggr|_{r=r_0}.
\end{equation}
where $x \equiv \sqrt{g_{00}(r_0)} = e^{A(r_0)}B(r_0)$ parameterizes the location of the boundary. 
Note that the counterterm action density is completely
local and that only even powers of $\chi$ appear (this is necessitated by the fact that only even 
powers of $z$ appear in the near-boundary series for $\Atilde$, $B$, and $\alpha$).
By tuning the coefficients $c_{i = 1, \dotsc, 11}$ appropriately,
the regulated action $S_\text{reg} \equiv S_\text{bulk} + S_\text{bdy} + S_\text{ctr}$ 
may remain finite upon taking the limit $r_0 \to \infty$.

The on-shell bulk action, with the Gibbons-Hawking term evaluates to
\begin{equation}
S_\text{bulk} = \frac{\beta V}{8\pi G_5}\biggl[e^{3A}\frac{d}{dr}(e^A B)\biggr]^{r_0}_0=\frac{\beta V}{8\pi G_5}\,\Delta - S,
\end{equation}
where, using eqs.~\eqref{dict:temperature} and \eqref{dict:entropy}, the term obtained from the lower limit
is just the entropy $S$.
The quantity $\Delta$ diverges in the limit $r_0 \to \infty$. Switching to the $z$ radial 
coordinate (so that the hypersurface sits at $z = z_0 = e^{-r_0/L}$), we see that the leading
singular behavior of $\Delta$ arises from $e^{4A(z_0)} \sim z_0^{-4}$. 
The counterterm coefficients are
chosen to remove the sensitivity of $\Delta$ on the cutoff in the limit $z_0 \to 0$. A 
straightforward expansion in powers of $z_0^{-1}$ and $\ln z_0$ leads to
\begin{equation}
c_1 = \tfrac{3}{2L}, \quad c_3 = 0, \quad c_4 = \tfrac{1}{L}, \quad
c_5 = \tfrac{6}{L}, \quad c_6 = 0, \quad c_8 = -\tfrac{3}{L}, \quad
c_{10} = -\tfrac{4}{3L}.
\end{equation}
The coefficient $c_{11}$ multiplies a finite counterterm and corresponds to a shift in the subtraction 
scheme. It may be fixed unambiguously as follows.
The renormalized supergravity action is identified with the gauge theory free energy divided by
the temperature, $S_\text{ren} \equiv \lim_{z_0 \to 0} S_\text{reg} = \beta F$. Using 
$F = E - TS$, the energy is given by $E = \frac{V}{8\pi G_5}\lim_{z_0 \to 0}\Delta$. At zero
temperature, the energy must vanish since the ground state is supersymmetric. For $\Delta$
to vanish in this limit, $c_{11}$ must equal $\tfrac{1}{3L}$. Ultimately,
\begin{equation}
\lim_{z_0 \to 0}\Delta = -\tfrac{3}{L}B_{4,0} - \tfrac{2}{L}\chi_{0,0}\bigl[
\chi_{2,0} - (2\alpha_{0,0}\chi_{0,0}-\third\chi_{0,0}^3)\bigr].
\end{equation}

Converting from $1/L$ to $T$, and using eq.~\eqref{nontrivial}, the free energy density can be
written as
\begin{equation}
\label{free_energy_sugra}
F/\vol = -\frac{\pi^2}{8}\Nc^2 T^4 \, 
\Bigl(\frac{2}{B_h}\Bigr)^3
\biggl[
1 - \frac{2}{B_{4,0}}
\Bigl(\chi_{2,0}\,\chi_{0,0} + \third \chi_{0,0}^4 - 2\alpha_{0,0}\,\chi_{0,0}^2\Bigr)
\biggr].
\end{equation}
The overall factor $- \tfrac{\pi^2}{8}\Nc^2 T^4$ is the free energy density of 
$\Nfour$ SYM. The remainder of the expression constitutes the scaling function $f$ in 
eq.~\eqref{free_energy_gauge}.

In ideal fluid hydrodynamics, the energy-momentum tensor $T^{\mu\nu}$ can be expressed in terms of the 
local fluid four-velocity $u^\mu(\mathbf{x})$ as 
$T^{\mu\nu} = (\epsilon + p)u^\mu u^\nu + p\, g^{\mu\nu}$, 
where the energy density $\epsilon \equiv E/\vol$ and the pressure $p = -\p F/\p V$. 
The trace is given by
\begin{equation}
\begin{split}
-\vev{T^\mu_{~\mu}} &= \epsilon - 3p = (E+3F)/\vol \\
&= \Nc^2 T^4 \,
\frac{\pi^2}{B_{4,0}} \Bigl(\frac{2}{B_h}\Bigr)^3
\Bigl(\chi_{2,0}\chi_{0,0} + \third \chi_{0,0}^4 - 2\alpha_{0,0}\chi_{0,0}^2\Bigr).
\end{split}
\end{equation}
In a conformal theory, $\epsilon = 3p$. The factor following $\Nc^2 T^4$ is the scaling function
$\omega$ in eq.~\eqref{EoS}.

We now describe how to obtain supergravity formulas for the thermal expectation values of the 
operators $\O_2$ and $\O_3$ using methods described in refs.~\cite{deHaro, Bianchi}.
These expectation values are given by functional derivatives of the renormalized 
supergravity action with respect to the couplings $m^2$ and $m$ (which, for this discussion, should
be thought of as independent couplings),
\begin{subequations}
\begin{align}
\vev{\O_2} &\equiv (\beta V)^{-1} \frac{\delta S_\text{ren}}{\delta m^2} 
= \lim_{r_0 \to \infty}\frac{1}{\beta V}\frac{\delta S_\text{reg}}{\delta \alpha(r_0)} 
\frac{\delta \alpha(r_0)}{\delta m^2}, \\
\vev{\O_3} &\equiv (\beta V)^{-1} \frac{\delta S_\text{ren}}{\delta m} 
= \lim_{r_0 \to \infty}\frac{1}{\beta V}\frac{\delta S_\text{reg}}{\delta \chi(r_0)} 
\frac{\delta \chi(r_0)}{\delta m}.
\end{align}
\end{subequations}
Let us evaluate the first functional derivative in each of the above expressions. Computing the
scalar functional derivatives for $S_\text{ctr}$ is straightforward with eq.~\eqref{counterterm},
so let us focus on $S_\text{bulk}$. We have
\begin{equation}
\begin{split}
S_\text{bulk} &= -\frac{1}{4\pi G_5}\int d^5x\, \sqrt{g} 
\Bigl(\fourth R + \mathcal{L}_\text{matter}\Bigr) \\
&= \frac{1}{4\pi G_5}\int d^5x\, \sqrt{g}
\Bigl[5\dot{A}^2 + \tfrac{5}{2}\dot{A}\dot{B}/B + 2\ddot{A} + \half\ddot{B}/B
+ 3\dot{\alpha}^2 + \dot{\chi}^2 + V(\alpha,\chi)\Bigr] \\
&= \frac{1}{4\pi G_5}\int d^5x\, \sqrt{g}
\Bigl[\half(\dot{K} + K^{\mu\nu}K_{\mu\nu}) + \tfrac{3}{2}(2\dot{A}^2 + \dot{A}\dot{B}/B) 
+ 3\dot{\alpha}^2 + \dot{\chi}^2 + V(\alpha,\chi)\Bigr] \\
&= \frac{1}{4\pi G_5}\int d^5x\, \sqrt{g}
\Bigl[\half(-K^2 + K^{\mu\nu}K_{\mu\nu}) + 6\dot{\alpha}^2 + 2\dot{\chi}^2\Bigr]
+ \frac{1}{8\pi G_5} \int d^4x\, \sqrt{g} K\Bigr|_{r=r_0}.
\end{split}
\end{equation}
The functional derivatives of the action are to be
evaluated at $r_0$. They should be viewed as the canonical momenta conjugate 
to $\alpha$ and $\chi$. One finds,
\begin{subequations}
\begin{align}
\frac{\delta S_\text{reg}}{\delta \alpha(r_0)} &= \frac{\beta V}{4\pi G_5}\, e^{4A}B 
\Bigl(6\dot{\alpha} + 2c_5\alpha + 2c_8\frac{\alpha}{\ln x}\Bigr)\biggr|_{r=r_0}, \\
\frac{\delta S_\text{reg}}{\delta \chi(r_0)} &= \frac{\beta V}{4\pi G_5}\, e^{4A}B
\Bigl(2\dot{\chi} + 2c_4\chi + 4c_{10}\chi^3\ln x + 4c_{11}\chi^3\Bigr)\biggr|_{r=r_0}.
\end{align}
\end{subequations}
Continuing with the calculation, write
\begin{subequations}
\begin{align}
\frac{\delta \alpha(z_0)}{\delta m^2} &= 
\frac{\delta \alpha_{0,1}}{\delta m^2} \frac{\delta \alpha(z_0)}{\delta\alpha_{0,1}} =
\tfrac{L^2}{6} z_0^2\bigl[\ln z_0 + O(z_0^2\ln^2 z_0)\bigr], \\
\frac{\delta \chi(z_0)}{\delta m} &= 
\frac{\delta \chi_{0,0}}{\delta m} \frac{\delta \chi(z_0)}{\delta\chi_{0,0}} = 
\tfrac{L}{2} z_0\bigl[1 + O(z_0^2\ln z_0)\bigr].
\end{align}
\end{subequations}
Putting it all together and sending the cutoff to zero (it is convenient to do this in the $z$ 
coordinate) yields
\begin{align}
\vev{\O_2} &= \frac{\Nc^2}{2\pi^2 L^2}\,\alpha_{0,0}, \qquad
\vev{\O_3} = -\frac{\Nc^2}{\pi^2 L^3}\,\bigl(\chi_{2,0} + \third \chi_{0,0}^3\bigr).
\end{align}
As expected, the subleading coefficients of the scalar fields in 5D supergravity are
directly related to the expectation values of the 4d gauge theory operators to which the leading 
coefficients couple. Lastly, one may check that the expectation values we have computed satisfy the 
conformal Ward identity,
\begin{equation}
\label{Ward}
-\vev{T^\mu_{~\mu}} = 2m^2\vev{\O_2} + m\vev{\O_3}.
\end{equation}
At zero temperature, one may insert the value of $\chi_{2,0}$ found in eq.~\eqref{subleading} and confirm that 
$\lim_{T\to0}\>\vev{T^\mu_{~\mu}} = 0$.

\subsection*{Endpoints of thermal scalar flows}
\label{app:thermalflows}

\begin{figure}[t]
    \centerline{\includegraphics[scale=0.85]{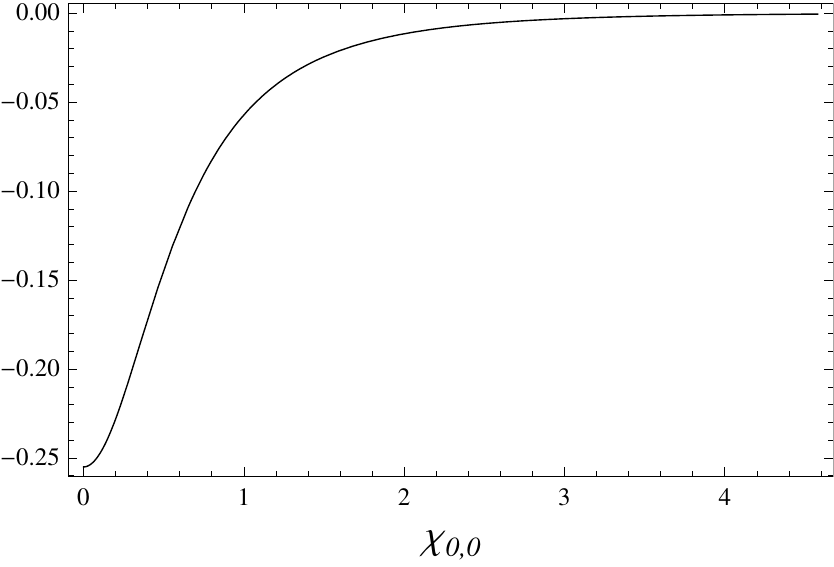}\qquad
    \includegraphics[scale=0.85]{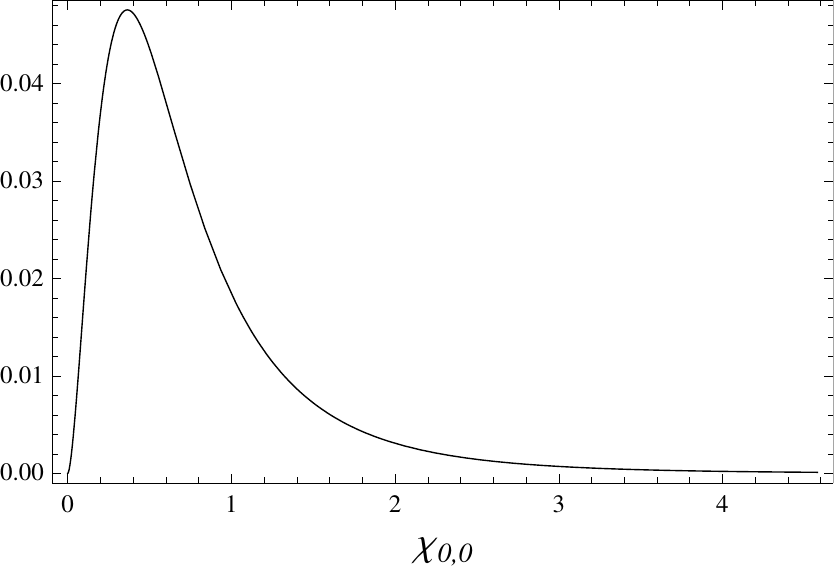}}
    \vspace*{-10pt}
    \caption{Deviation of endpoints of thermal flows from supersymmetric flow equations.
             Left: a plot of $e^{A_h} - 2\chi_{0,0} e^{2\alpha_h}/\sinh 2\chi_h$ vs. $\chi_{0,0}$.
             This measures how well eq.~\eqref{susyflow1} is satisfied at the horizon.
             Right: a plot of 
             $e^{6\alpha_h} - [\cosh 2\chi_h + (\gamma + \ln\tanh\chi_h)\sinh^2(2\chi_h)]$ vs.
             $\chi_{0,0}$.
             This measures how well eq.~\eqref{susyflow2} is satisfied at the horizon.
}
    \label{fig:thermalflows}
\end{figure}

Figure~\ref{fig:flows} showed the trend that the thermal solutions for the scalars $\alpha$
and $\chi$ became successively better approximations to the zero temperature enhan\c{c}on solution
as $m/T \to \infty$. In figure~\ref{fig:thermalflows} we provide further evidence for this assertion
by plotting the residuals to eq.~\eqref{susyflow}, evaluated at the horizon,
for our numerical backgrounds.
On the left is a plot of $e^{A_h} - 2\chi_{0,0} \, e^{2\alpha_h}/\sinh 2\chi_h$,
and on the right
a plot of $e^{6\alpha_h} - [\cosh 2\chi_h + (\gamma + \ln\tanh\chi_h)\sinh^2(2\chi_h)]$ for 
$\gamma = 0$. In each plot the deviation from zero vanishes as we approach larger values of 
$\chi_{0,0}$, which corresponds to lower temperatures.

\subsection*{First law of thermodynamics}

The first law of thermodynamics, $dE = TdS$,
is equivalent to the statement that $dF/dT = -S$.
Let us express this constraint in terms of supergravity parameters.
For temporary convenience define 
$\Omega \equiv \chi_{2,0}\,\chi_{0,0} + \third\chi_{0,0}^4 - 2\alpha_{0,0}\,\chi_{0,0}^2$. 
Differentiating the formula for the free energy density in eq.~\eqref{free_energy_sugra} gives
\begin{equation}
\frac{dF}{dT} = -S \> \biggl\{
1 - \frac{2\Omega}{B_{4,0}} + \fourth B_h^3\, T \frac{d}{dT}\,\biggl(\frac{1-2\Omega/B_{4,0}}{B_h^3}\biggr)
\biggr\}.
\end{equation}
The entire term inside the braces must equal 1.
Note that the temperature derivative acts on 
objects which are functions of $\chi_{0,0}$.
Using the chain rule,
$T\frac{d}{dT} = T\, \frac{d\chi_{0,0}}{dT}\,\frac{d}{d\chi_{0,0}}$,
we need
\begin{equation}
T\,\frac{d\chi_{0,0}}{dT} = \left[{A_h' + B_h'/B_h - 1/\chi_{0,0}}\right]^{-1}.
\end{equation}
This relation comes from differentiating eq.~\eqref{mbeta}, with primes denoting
$d/d\chi_{0,0}$.
After straightforward manipulation, the first law condition becomes%
\footnote{%
This simple form is based on the assumption that the relation between $T$ and $\chi_{00}$ is monotonic. 
We have verified numerically that $B_h'/B_h +A_h' \neq 1/\chi_{0,0}$. Notice that $d\chi_{0,0}/dT$ is 
always nonzero since $B_h$ is always positive; this follows from the positivity of the entropy.
}
\begin{equation}
\label{firstlaw}
\half\chi_{2,0}'\chi_{0,0} - \tfrac{3}{2}\chi_{2,0} 
- \alpha_{0,0}'\chi_{0,0}^2 + 2\alpha_{0,0}\chi_{0,0}
+ \tfrac{3}{4}B_{4,0}' - 3A_h'B_{4,0} = 0. 
\end{equation}
This is a 1st order ODE with respect to the parameter $\chi_{0,0}$,
involving four of the six independent horizon and boundary data.
Therefore, it
is a nontrivial constraint which serves as a robust check on our numerics \cite{BDKL}.

\begin{figure}[t]
    \centerline{\includegraphics[scale=1.1]{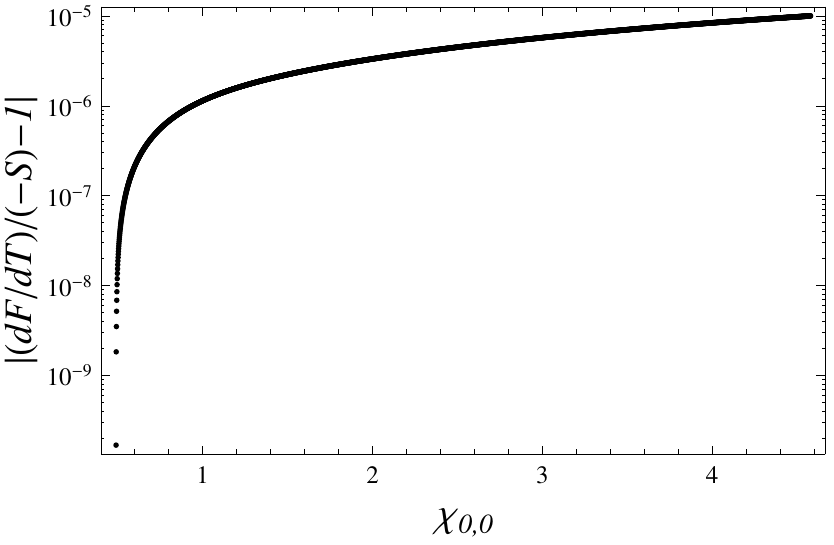}}
    \vspace*{-10pt}
    \caption{Deviation from first law of thermodynamics. The left hand side of 
    eq.~\eqref{firstlaw} is plotted versus $\chi_{0,0}$. The deviation may be attributed to 
    discretization error.}
    \label{fig:firstlaw}
\end{figure}

The left hand side of eq.~\eqref{firstlaw} is plotted in figure~\ref{fig:firstlaw}.
While no numerical computation with finite precision numbers can ever give a result which is 
exactly zero, the scale on our plot indicates that the first law is satisfied accurately to within 
$10^{-5}$ for the largest $\chi_{0,0}$ (or lowest temperatures) explored. 
The deviation from zero is due to discretization effects.%
\footnote
{%
To demonstrate that the deviation is a discretization effect,
we tried using a better approximation for the derivative.
With the fourth order approximation,
$$
\frac{\mathsf{F}(\chi_{0,0} {-} 2h) - 8\mathsf{F}(\chi_{0,0} {-} h) + 8\mathsf{F}(\chi_{0,0} {+} h)
- \mathsf{F}(\chi_{0,0} {+} 2h)}{12h} 
= \mathsf{F}'(\chi_{0,0}) + O(h^4).
$$
we found that the deviation from the first law of thermodynamics was, at worst, of order $10^{-9}$ for 
$\chi_{0,0} \gtrsim 3.7$ and even smaller for lower values of $\chi_{0,0}$.
}

\section{Numerical procedure}
\label{app:numerics}

We used a \emph{shooting technique} to determine the independent boundary and 
horizon data \eqref{data} as functions of $\chi_{0,0}$. 
The algorithm is as follows.
For a given value of $\chi_{0,0}$, start with a trial set of independent
boundary and horizon data,
\begin{equation}
\mathbf{X}(\chi_{0,0}) = (B_{4,0}, \alpha_{0,0}, \chi_{2,0}, A_h, \alpha_h, \chi_h)(\chi_{0,0}).
\end{equation}
The remaining parameters are fixed by $\alpha_{0,1} = \tfrac{2}{3}\chi_{0,0}^2$ and 
$B_h = -4B_{4,0}e^{-4A_h}$.
One may now construct a near-horizon series solution using eq.~\eqref{nearhorsol}. Use this
series to evaluate
\begin{equation}
\mathbf{V}(z) \equiv (B, \Atilde, \alpha, \alpha', \chi, \chi')(z)
\end{equation}
at some $z_\text{max}$ close to the horizon. The output is some vector 
$\mathbf{V}(z_\text{max})|_\text{series}$ which provides initial conditions for the equations of 
motion given in eq.~\eqref{ODEs}.
Integrate this system from $z_\text{max}$ down to a point $z_*$ 
in the middle of the bulk. Evaluating the fields and their derivatives at this point produces 
$\mathbf{V}(z_*)|_{\text{hor}\to\text{bulk}}$. Now repeat this process from 
the other direction. Construct the near-boundary series solution using eq.~\eqref{nearbdysol} and 
use it to evaluate $\mathbf{V}(z_\text{min})|_\text{series}$, where $z_\text{min}$ is close to 
the boundary. This provides the initial conditions needed to integrate the system from 
$z_\text{min}$ up to $z_*$. Evaluating the solution at the last point gives 
$\mathbf{V}(z_*)|_{\text{bdy}\to\text{bulk}}$. Finally,
take the difference
\begin{equation}
\mathbf{M} \equiv 
\mathbf{V}(z_*)\bigr|_{\text{bdy}\to\text{bulk}}-\mathbf{V}(z_*)\bigr|_{\text{hor}\to\text{bulk}}.
\end{equation}
This is called the `mismatch vector' \cite{Aharony}. 

The system of ODEs requires only the values of $B$ and 
$\Atilde$, and the values and slopes of $\alpha$ and $\chi$, at a single point to completely 
fix the behavior of the metric and scalar fields throughout the spacetime. If the system is 
reexpressed as a set of six 1st order ODEs, then a nonzero mismatch vector means that one or more
fields are discontinuous across $z_*$. This implies that our initial choice of $\mathbf{X}$
produced inconsistent initial conditions at the horizon and boundary. The correct choice of 
$\mathbf{X}$ must lead to $\mathbf{M} = 0$. By thinking of $\mathbf{M}(\mathbf{X})$
as a vector-valued function, the problem becomes that of root finding in six dimensions. 
We apply the Newton-Raphson method (see, {e.g.}, ref.~\cite{Recipes}). 
It works by a generalization of
the familiar one-dimensional method of tracking tangent lines. For a guess $\mathbf{X}$, compute 
the Jacobian $\mathbf{J}$ of partial derivatives of the mismatch vector
($J_{ij} \equiv \p_j M_i$). Then form the
new guess $\mathbf{X}_\text{new} = \mathbf{X} - \mathbf{J}^{-1}\mathbf{M}$. Iterate to
produce a sequence of $\mathbf{X}$'s that converge to the true root.

\begin{figure}[t]
\centerline{\includegraphics[scale=0.9]{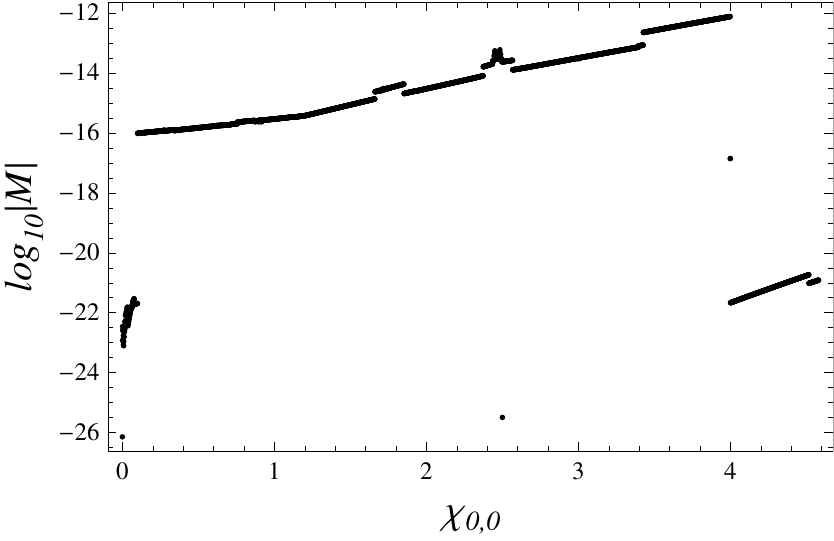}}
\vspace*{-10pt}
\caption{Manhattan norm of mismatch vector, plotted as a function of $\chi_{0,0}$.}
\label{fig:norm}
\end{figure}

We implemented this algorithm in {\it Mathematica 7.0}. We let $z_\text{min} \equiv d$ and 
$z_\text{max} \equiv 1-d$ with $d = 1/1000$.
We perform a series expansion of the solution close to the boundary,
and close to the horizon.
Near the horizon the series is evaluated up to and including terms of
order $(z{-}1)^6$ for each of the four supergravity fields. Near the boundary the series is 
evaluated up to and including terms of order $z^{10}\ln^4 z$ for $B$,
$z^{10}\ln^6 z$ for $\Atilde$, 
$z^{10}\ln^6 z$ for $\alpha$, and $z^9\ln^5 z$ for $\chi$. The equations of motion were integrated
using {\tt NDSolve} with {\tt WorkingPrecision} set to 40 digits, {\tt MaxSteps} set to $\infty$,
and {\tt PrecisionGoal} and {\tt AccuracyGoal} each
set to 20 digits. The matching was performed at $z_* = 1/2$. For Newton's method we found that a 
step size of $1/1000000$ was adequate to compute a forward finite difference approximation to the 
Jacobian. We iterated until the Manhattan norm of the mismatch vector, 
$|\mathbf{M}| \equiv \sum_{i=1}^6 |M_i|$, was below a threshold of $10^{-7}$.
In practice, for a sufficiently good starting guess for $\mathbf{X}$, only one or two Newton steps 
were needed to obtain an incredibly small norm. The data in figure~\ref{fig:norm} shows how low our
norms became after iterating Newton's method more than a couple times.

For good numerical performance,
it is important to independently integrate \emph{inward} from both the boundary and horizon,
since $z = 0$ and 1 are regular singular points of the ODEs. 
To see this, observe that the field equations 
for $\alpha$ and $\chi$ reduce to decoupled Euler differential equations.%
\footnote{%
Near $z = 0$ they become $\alpha'' - (3/z)\,\alpha' + (4/z^2)\,\alpha = 0$ and 
$\chi'' - (3/z)\,\chi' + (3/z^2)\,\chi = 0$. Near $z = 1$ they become $\alpha'' + \alpha'/(z{-}1) = 0$ 
and $\chi'' + \chi'/(z{-}1) = 0$.
}
In all instances, the roots of the indicial equation differ by an integer, and this means that one 
of the linearly independent solutions has a logarithmic singularity.
Any singular behavior
is unwanted near the horizon where the fields should be completely regular.
Therefore, we seek a 
solution in which the coefficient of $\ln(z{-}1)$ vanishes as we approach $z = 1$ from below. 
However, integrating toward this endpoint is a bad idea since the logarithm grows, and the solution
we do want stays constant. Numerical error has the effect of putting noise into the coefficients 
of the linearly independent solutions, so if the unphysical solution grows faster than the 
good solution, the relative size of numerical error becomes worse the closer one approaches to the horizon.
It is also unhelpful to integrate toward $z = 0$ from above since both linearly independent 
solutions vanish and a numerically ill-behaved limiting procedure would be
needed to isolate the unphysical solution. These
practical issues are alleviated by starting from the correct solutions at the horizon and boundary,
and integrating into the bulk where no singular points exist. Although our choice for
the ``middle of the bulk" was arbitrary, we verified insensitivity to its
choice a posteriori.

A crucial question that must be addressed in any numerical computation is: how many significant 
digits in the final result can be trusted? 
We made extensive use of {\it Mathematica}'s ability to represent 
arbitrary precision numbers. Testing showed that integrating with 40 digit working precision was 
more than sufficient to guarantee that the mismatch vector's individual components did not change 
to within $10^{-7}$.
In particular, our results are also 
sensitive to the value of $d$ and the truncation order $n$ of the series. The sense in which 
$d = 1/1000$ is small must be examined in light of how large $n$ is. There is a simple relation 
between the two: making $d$ smaller is akin to making $n$ bigger since each additional order in a 
series is roughly suppressed by a factor of $d$ compared to the previous order. %
We kept $d = 1/1000$ fixed and decreased $n$ to the next nontrivial order for both series 
solutions. 
Solving for the roots as before 
(stopping when a threshold of $10^{-7}$ for the norm of the mismatch vector was crossed), 
we found that the new roots were always identical to the original roots in the first 7 
significant digits. 

To generate good initial guesses for the roots,
we linearly extrapolated the last two known roots along
the $\chi_{0,0}$-axis. We found 4581 roots spaced at intervals of 0.001 from $\chi_{0,0} = 0$ to
$4.58$. For the case $\chi_{0,0} = 0$, the exact root is 
$\mathbf{X}(0) = (-2, 0, 0, \ln \sqrt{2}, 2, 0, 0)$, which follows from the AdS-Schwarzschild 
solution.%
\footnote{%
A standard form of the AdS-Schwarzschild metric (in units where $L = 1$) is
$
ds^2 = \rho^2[-f(\rho) \, dt^2 + d\mathbf{x}^2] + \rho^{-2} {d\rho^2}/f(\rho)
$,
where $f(\rho) = 1- (\rho_h/\rho)^4$ and $\rho_h \equiv \pi T$. This can be rewritten in the 
form of eq.~\eqref{metric} by defining $\rho = \rho_h\sqrt{\cosh 2r}$. The horizon
is located at $r = 0$. A further change of variables to $z = e^{-r}$ puts the metric into the form of 
eq.~\eqref{metricz}, from
which we find that $B(z) = (1-z^4)/(1+z^4)$. A Taylor expansion around the origin gives 
$B(z) = 1 - 2z^4 + O(z^8)$ from which we read off $B_{4,0} = -2$. Also, $B_h = -B'(1) = 2$.
From 
eq.~\eqref{nontrivial} we find that $A_h = \ln\sqrt{2}$. Lastly, the scalar fields are identically
zero in this background.
}
Naive guess-and-check was employed to find the first few roots for $\chi_{0,0}>0$.
Although Newton's method can diverge rapidly if a poor starting guess is made, in practice, we
found it to be quite forgiving. Slightly fancier methods involving
backtracking were used (quite infrequently) to search for roots when the linear extrapolation 
failed to produce a reasonable guess \cite{Recipes}.

\section{Fluctuation analysis}
\label{app:flucs}
\subsection*{Tensor channel}
\label{app:tensor}

For numerical integration it is convenient to consider the system
\begin{equation}
\psi' + (4\Atilde' + B'/B - 3/z)\psi + \kappa^2 e^{-2\Atilde}\phi = 0, \qquad
\phi' = \psi.
\end{equation}
Since $z = 0, 1$ are singular points, we use series solutions to provide initial conditions
at $z_\text{min}$ and $z_\text{max}$, then integrate inward to a matching point $z_*$. The
regular near-horizon solution may be written in the form
\begin{equation}
\phi(z) = c_0\biggl[1 + \sum_{n \geq 1} c_n (z{-}1)^n\biggr].
\end{equation}
We evaluate the coefficients up to $c_6$. 
The normalizable near-boundary solution has the form
\begin{equation}
\phi(z) = b_{0,0}\,z^4\biggl[ 1+ b_{0,1}\ln z + \sum_{n \geq 1} z^n \sum_{k = 0}^n b_{n,k} 
(\ln z)^k \biggr],
\end{equation}
where we compute the coefficients up to $b_{8,4}$.

Our algorithm for solving the small fluctuation equations is
similar to that used for the background and discussed in
Appendix \ref{app:numerics}. For a given 
$\chi_{0,0}$, the near-horizon amplitude $c_0$ is arbitrarily set to 1
and a guess is made for the eigenvalue $\kappa$ and the near-boundary
amplitude $b_{0,0}$. 
Next $\phi$ and its derivative $\psi$ are evaluated with the series at $z_\text{min}$ and $z_\text{max}$, 
which are then fed as initial conditions into {\tt NDSolve}. Using interpolations of the 
numerically-generated background fields, the system of ODEs is integrated inward to the matching
point $z_*$. We used machine precision settings for {\tt NDSolve}. The mismatch vector for 
$(\phi(z_*), \psi(z_*))$ is computed and Newton's method in two dimensions is iterated
(the Jacobian was calculated with a step size of $1/1000$). Using 5 Newton steps, we obtained 
Manhattan norms typically several orders of magnitude below $10^{-7}$. This yields an estimate 
for the root $\kappa$ in units of $1/L$.
To find $\kappa$ in units of $\pi T$, we multiply the root by $2/(e^{A_h}B_h)$.

\subsection*{Vector channel}
\label{app:vector}

The fluctuation equations in the vector channel are
\begin{equation}
\psi' + (4\Atilde' - B'/B - 3/z)\psi + \kappa^2 e^{-2\Atilde}\phi = 0, \qquad
\phi' = \psi.
\end{equation}
The regular near-horizon series solution is
\begin{equation}
\phi(z) = c_0 \, (z{-}1)^2 \biggl[1 + \sum_{n \geq 1} c_n (z{-}1)^n\biggr],
\end{equation}
for which we compute the coefficients up to $c_5$.
The normalizable near-boundary series solution is
\begin{equation}
\phi(z) = b_{0,0}\, z^4\biggl[1 + b_{0,1}\ln z + \sum_{n \geq 1} z^n \sum_{k = 0}^n b_{n,k}
(\ln z)^k \biggr],
\end{equation}
for which we compute the coefficients up to eighth order.
The numerical algorithm is identical to that for the tensor channel.

\subsection*{Scalar channel}
\label{app:scalar}

{\it Derivation of gauge-invariant equations for helicity zero}
\medskip

We outline the necessary steps to obtain eqs.~\eqref{Zac}.
The $3z$, $00$, $33$, $zz$, and sum of the $11$ and $22$ Einstein equations give (in order)
\begin{subequations}
\begin{align}
& \label{3z_eq}
H_{00}' + H_+' + b' H_{00} = -8(3\alpha'_\text{cl}\alphafluc + \chi'_\text{cl}\chifluc), \\
& \label{00_eq}
H_{00}'' + (5A'+2b'+1/z)H_{00}' + (A'+b')(H_+' + H_{33}') 
+ \kappa^2\frac{e^{-2A}}{z^2}H_{00} \\
& \quad \nonumber
= -\frac{8(V_\alpha\alphafluc + V_\chi\chifluc)}{3z^2}, \\
& \label{33_eq}
H_{33}'' + (5A'+b'+1/z)H_{33}' + A'(H_{00}' + H_+') 
+ \kappa^2\frac{e^{-2A}}{z^2} (H_{00}+H_+) \\
& \quad \nonumber
= -\frac{8(V_\alpha\alphafluc + V_\chi\chifluc)}{3z^2}, \\
& \label{zz_eq}
H_{00}'' + H_+'' + H_{33}'' + (2A'+2b'+1/z)H_{00}' + (2A'+1/z)(H_+' + H_{33}') \\
& \quad \nonumber
= - 16(3\alpha'_\text{cl}\alphafluc' + \chi'_\text{cl}\chifluc')
-\frac{8(V_\alpha\alphafluc + V_\chi\chifluc)}{3z^2}, \\
& \label{plus_eq}
H_+'' + (6A'+b'+1/z)H_+' + 2A'(H_{00}' + H_{33}')
+ \kappa^2\frac{e^{-2A}}{z^2} H_+
= -\frac{16(V_\alpha\alphafluc + V_\chi\chifluc)}{3z^2}.
\end{align}
\end{subequations}
The scalar field equations are
\begin{subequations}
\begin{align}
& \label{alpha_eq}
\alphafluc'' + (4A'+b'+1/z)\alphafluc' + \half\alpha'_\text{cl}(H_{00}' + H_+' + H_{33}')
+ \kappa^2\frac{e^{-2A}}{z^2}\alphafluc
= \frac{V_{\alpha\alpha}\alphafluc + V_{\alpha\chi}\chifluc}{6z^2}, \\
& \label{chi_eq}
\chifluc'' + (4A'+b'+1/z)\chifluc' + \half\chi'_\text{cl}(H_{00}' + H_+' + H_{33}')
+ \kappa^2\frac{e^{-2A}}{z^2}\chifluc
= \frac{V_{\chi\chi}\chifluc + V_{\alpha\chi}\alphafluc}{2z^2}.
\end{align}
\end{subequations}
Hence, there are seven coupled ODEs in axial gauge.%

In simplifying the above system it is helpful to use the background equations.
These relations make it possible to express second (or higher order) derivatives of the 
background fields in terms of their first derivatives. 
The first step is to recognize that $H_{33}$ does not appear in any of the ODEs, only its
derivatives, so one more equation is first order in $H_{33}'$ and can be considered a constraint. This allows one to drop one of the 2nd order equations, say eq.~\eqref{zz_eq}. The second step is to eliminate $H_{33}'$ from the remaining equations. To do this, sum 
eqs.~\eqref{00_eq} and \eqref{plus_eq}, then solve for $H_{33}'$. 
As the other independent equation, keep eq.~\eqref{00_eq}. Now plug in for $H_{33}'$ in all four equations.
The third step is to use the residual gauge invariance that preserves the axial gauge, $h_{\mu z} = 0$, to eliminate $H_+$.
In axial gauge, eq.~\eqref{gauge}, 
there is a residual gauge invariance parameterized by a vector field 
$\eta_\mu \equiv \eta_\mu(z) e^{-\kappa x^3}$ which obeys 
$-\nabla_\mu\eta_z - \nabla_z\eta_\mu = 0$. As a contravariant vector, 
\begin{equation}
\eta^\mu(z) = (C_0,\, C_1,\, C_2,\, C_3 + C_z\, \kappa \int^z \! dz'\, e^{-2A(z')}/z',\, C_z z),
\end{equation}
for arbitrary constants $C_\mu$.%
\footnote{%
If $C_z = 0$, then the residual gauge freedom simply corresponds to
translations in the 4d space $(\tau, \mathbf{x})$.
}
While this preserves $h_{\mu z} = 0$, consider what happens to the other fluctuation components. 
In particular,
\begin{subequations}
\begin{align}
\delta h_{00}(z) &= -2C_z\, z\, B^2 e^{2A} (A' + B'/B), \\
\delta h_+(z) &= -4C_z\, z\, A' e^{2A}, \\
\delta h_{33}(z) &= 2e^{2A}\Bigl[ C_3\kappa + C_z\kappa^2 \int^z \! dz'\, e^{-2A(z')}/z'
-C_z\, z\, A'\Bigr], \\
\delta \alphafluc(z) &= -C_z \, z \, \alpha'_\text{cl}, \\
\delta \chifluc(z) &= -C_z\,  z\,  \chi'_\text{cl}.
\end{align}
\end{subequations}
One can easily verify that the fields defined in eq.~\eqref{gauge_invts} are
gauge invariant. 
Finally, some algebra and judicious rearrangement are needed to obtain eqs.~\eqref{Zac}. 

\medskip
\noindent {\it Near-horizon solution}
\medskip

The regular near-horizon series solutions are
\begin{subequations}
\begin{align}
\Ztilde(z) &= \Ztilde_0 + \sum_{n \geq 1}\Ztilde_n (z{-}1)^n, \\
\atilde(z) &= \atilde_0 + \sum_{n \geq 1}\atilde_n (z{-}1)^n, \\
\ctilde(z) &= \ctilde_0 + \sum_{n \geq 1}\ctilde_n (z{-}1)^n.
\end{align}
\end{subequations}
It is straightforward to determine the coefficients by solving eq.~\eqref{Zac} order-by-order
in $z-1$. We did this up to fourth order. We find that $\Ztilde_1 = -2\Ztilde_0$, 
$\atilde_1 = -2\atilde_0$, and $\ctilde_1 = -2\ctilde_0$. Higher coefficients are lengthy and we
do not write them down.

\medskip
\noindent {\it Near-boundary solution}
\medskip

The normalizable near-boundary series solutions are
\begin{subequations}
\begin{align}
\Ztilde(z) &= z\biggl[\Ztilde_{0,0} + \Ztilde_{0,1}\ln z 
+ \sum_{n \geq 1} z^n \sum_{k=0}^n \Ztilde_{n,k} (\ln z)^k\biggr], \\
\atilde(z) &= \atilde_{0,0} + \atilde_{0,1}\ln z
+ \sum_{n \geq 1} z^n \sum_{k=0}^n \atilde_{n,k} (\ln z)^k, \\
\ctilde(z) &= z\biggl[\ctilde_{0,0} + \ctilde_{0,1}\ln z
+ \sum_{n \geq 1} z^n \sum_{k=0}^n \ctilde_{n,k} (\ln z)^k\biggr],
\end{align}
\end{subequations}
where we compute the coefficients up to fifth order.
Odd coefficients up to and including $\atilde_{5,k}$ and $\ctilde_{5,k}$ are zero. Note that 
$\atilde_{0,1}$ is not fixed by eq.~\eqref{Zac} --- it is the leading coefficient of the
non-normalizable solution. We set it to zero by hand.

\bigskip
\noindent {\it Numerical procedure}
\medskip

The shooting method for the coupled set of ODEs cannot start from $z = 0$ or 1, as these are 
singular points. Therefore, it is necessary to evaluate the series solutions on the basis vectors 
at $z = z_\text{min}$ and $z_\text{max}$ to start the numerical integration. We used settings for 
{\tt NDSolve} and Newton's method identical to those for the tensor channel. However, in this case,
Newton's method only needs to be applied in one dimension, making the calculation much simpler.


\begin{thebibliography}{99}

\bibitem{ArnYaf}
  P.~B.~Arnold and L.~G.~Yaffe,
  {\it The non-Abelian Debye screening length beyond leading order},
  \prd{52}{1995}{7208},
  \hepph{9508280}.

\bibitem{Laine}
  M.~Laine and O.~Philipsen,
  {\it The non-perturbative QCD Debye mass from a Wilson line operator},
  \plb{459}{1999}{259},
  \heplat{9905004}.

\bibitem{Laer}
  E.~Laermann and O.~Philipsen,
  {\it Status of lattice QCD at finite temperature},
  \arnps{53}{2003}{163},
  \hepph{0303042}.

\bibitem{Tannen}
  M.~J.~Tannenbaum,
  {\it Recent results in relativistic heavy ion collisions: From ``a new state of matter" to 
  ``the perfect fluid"},
  \newjournal{Rept.\ Prog.\ Phys.\ }{RPPHA}{69}{2006}{2005},
  \nuclex{0603003}.

\bibitem{adscftA}
  J.~M.~Maldacena,
  {\it The large $N$ limit of superconformal field theories and supergravity,}
  \atmp{2}{1998}{231}
  [Int.\ J.\ Theor.\ Phys.\  {\bf 38} (1999) 1113],
  \hepth{9711200}.

\bibitem{adscftB}
  S.~S.~Gubser, I.~R.~Klebanov and A.~M.~Polyakov,
  {\it Gauge theory correlators from non-critical string theory,}
  \plb {428}{1998}{105},
  \hepth{9802109}.

\bibitem{adscftC}
  E.~Witten,
  {\it Anti-de Sitter space and holography,}
  \atmp{2}{1998}{253},
  \hepth{9802150}.

\bibitem{shear1}
  G.~Policastro, D.~T.~Son and A.~O.~Starinets,
  {\it The shear viscosity of strongly coupled $\Nfour$ supersymmetric Yang-Mills plasma,}
  \prl{87}{2001}{081601},
  \hepth{0104066}.

\bibitem{shear2}
  D.~T.~Son and A.~O.~Starinets,
  {\it Minkowski-space correlators in AdS/CFT correspondence: Recipe and applications,}
  \jhep{0209}{2002}{042},
  \hepth{0205051}.

\bibitem{boost}
  R.~A.~Janik and R.~B.~Peschanski,
  {\it Asymptotic perfect fluid dynamics as a consequence of AdS/CFT,}
  \prd{73}{2006}{045013},
  \hepth{0512162}.

\bibitem{fluidgrav}
  S.~Bhattacharyya, V.~E.~Hubeny, S.~Minwalla and M.~Rangamani,
  {\it Nonlinear fluid dynamics from gravity,}
  \jhep{0802}{2008}{045},
  \arXivid{0712.2456}.

\bibitem{drag1}
  C.~P.~Herzog, A.~Karch, P.~Kovtun, C.~Kozcaz and L.~G.~Yaffe,
  {\it Energy loss of a heavy quark moving through $\Nfour$ supersymmetric Yang-Mills plasma,}
  \jhep{0607}{2006}{013},
  \hepth{0605158}.

\bibitem{drag2}
  S.~S.~Gubser,
  {\it Drag force in AdS/CFT,}
  \prd{74}{2006}{126005},
  \hepth{0605182}.

\bibitem{drag3}
  J.~Casalderrey-Solana and D.~Teaney,
  {\it Heavy quark diffusion in strongly coupled $\Nfour$ Yang-Mills,}
  \prd{74}{2006}{085012},
  \hepph{0605199}.

\bibitem{quench}
  H.~Liu, K.~Rajagopal and U.~A.~Wiedemann,
  {\it Wilson loops in heavy ion collisions and their calculation in AdS/CFT,}
  \jhep{0703}{2007}{066},
  \hepph{0612168}.

\bibitem{thermalization}
  P.~M.~Chesler and L.~G.~Yaffe,
  {\it Horizon formation and far-from-equilibrium isotropization in supersymmetric Yang-Mills plasma,}
  \prl{102}{2009}{211601},
  \arXivid{0812.2053}.

\bibitem{Bak}
  D.~Bak, A.~Karch and L.~G.~Yaffe,
  {\it Debye screening in strongly coupled N=4 supersymmetric Yang-Mills plasma},
  \jhep{0708}{2007}{049},
  \arXivid{0705.0994}.

\bibitem{Amado}
  I.~Amado, C.~Hoyos-Badajoz, K.~Landsteiner and S.~Montero,
  {\it Absorption lengths in the holographic plasma},
  \jhep{0709}{2007}{057},
  \arXivid{0706.2750}.

\bibitem{Datta2}
  S.~Datta and S.~Gupta,
  {\it Dimensional reduction and screening masses in pure gauge theories at finite temperature,}
  \npb{534}{1998}{392},
  \heplat{9806034}.

\bibitem{DG}
  S.~Datta and S.~Gupta,
  {\it Does the QCD plasma contain propagating gluons?},
  \prd{67}{2003}{054503},
  \heplat{0208001}.

\bibitem{Hart}
  A.~Hart, M.~Laine and O.~Philipsen,
  {\it Static correlation lengths in QCD at high temperatures and finite densities},
  \npb{586}{2000}{443},
  \hepph{0004060}.

\bibitem{Donagi}
  R.~Donagi and E.~Witten,
  {\it Supersymmetric Yang-Mills theory and integrable systems},
  \npb{460}{1996}{299},
  \hepth{9510101}.

\bibitem{PW}
  K.~Pilch and N.~P.~Warner,
  {\it $\Ntwo$ supersymmetric RG flows and the IIB dilaton},
  \npb{594}{2001}{209},
  \hepth{0004063}.

\bibitem{BLthermo}
  A.~Buchel and J.~T.~Liu,
  {\it Thermodynamics of the $\Ntwo$\!* flow},
  \jhep{0311}{2003}{031},
  \hepth{0305064}.

\bibitem{Bhydro}
  A.~Buchel,
  {\it $\Ntwo$\!* hydrodynamics},
  \npb{708}{2005}{451},
  \hepth{0406200}.

\bibitem{BDKL}
  A.~Buchel, S.~Deakin, P.~Kerner and J.~T.~Liu,
  {\it Thermodynamics of the $\Ntwo$\!* strongly coupled plasma},
  \npb{784}{2007}{72},
  \hepth{0701142}.

\bibitem{Bviscosity}
  A.~Buchel,
  {\it Bulk viscosity of gauge theory plasma at strong coupling},
  \plb{663}{2008}{286},
  \arXivid{0708.3459}.

\bibitem{Bviscosity2}
  A.~Buchel and C.~Pagnutti,
  {\it Bulk viscosity of $\Ntwo$\!* plasma},
  \npb{816}{2009}{62},
  \arXivid{0812.3623}.

\bibitem{Hoyos}
  C.~Hoyos-Badajoz,
  {\it Drag and jet quenching of heavy quarks in a strongly coupled $\Ntwo$\!* plasma},
  \jhep{0909}{2009}{068},
  \arXivid{0907.5036}.

\bibitem{online}
    See \url{http://www.phys.washington.edu/users/lgy/N2starAppendices.pdf}.

\bibitem{Myers}
  S.~Kobayashi, D.~Mateos, S.~Matsuura, R.~C.~Myers and R.~M.~Thomson,
  {\it Holographic phase transitions at finite baryon density},
  \jhep{0702}{2007}{016},
  \hepth{0611099}.

\bibitem{BPP}
  A.~Buchel, A.~W.~Peet and J.~Polchinski,
  {\it Gauge dual and noncommutative extension of an $\Ntwo$ supergravity solution},
  \prd{63}{2001}{044009},
  \hepth{0008076}.

\bibitem{sugra1}
  M.~Gunaydin, L.~J.~Romans and N.~P.~Warner,
  {\it Gauged $\mathcal N=8$ supergravity in five-dimensions,}
  \plb{154}{1985}{268}.

\bibitem{sugra2}
  M.~Gunaydin, L.~J.~Romans and N.~P.~Warner,
  {\it Compact and noncompact gauged supergravity theories in five-dimensions,}
  \npb{272}{1986}{598}.

\bibitem{sugra3}
  M.~Pernici, K.~Pilch and P.~van Nieuwenhuizen,
  {\it Gauged $\mathcal N=8$, $D=5$ supergravity,}
  \npb{259}{1985}{460}.

\bibitem{KPW}
  A.~Khavaev, K.~Pilch and N.~P.~Warner,
  {\it New vacua of gauged N = 8 supergravity in five dimensions,}
  \plb{487}{2000}{14},
  \hepth{9812035}.

\bibitem{Evans}
  N.~J.~Evans, C.~V.~Johnson and M.~Petrini,
  {\it The enhancon and $\Ntwo$ gauge theory/gravity RG flows},
  \jhep{0010}{2000}{022},
  \hepth{0008081}.

\bibitem{Malda}
  J.~M.~Maldacena,
  {\it Lectures on AdS/CFT},
  \hepth{0309246}.

\bibitem{Carlos}
  C.~Hoyos-Badajoz,
  {\it Higher dimensional conformal field theories in the Coulomb branch,}
  \arXivid{1010.4438}.

\bibitem{noncftbranes}
  I.~Kanitscheider and K.~Skenderis,
  {\it Universal hydrodynamics of non-conformal branes},
  \jhep{0904}{2009}{062},
  \arXivid{0901.1487}.

\bibitem{SW1}
  N.~Seiberg and E.~Witten,
  {\it Monopole condensation and confinement in $\Ntwo$ supersymmetric Yang-Mills theory,}
  \npb{426}{1994}{19}
  [Erratum-ibid.\npb{430}{1994}{485}],
  \hepth{9407087}.

\bibitem{SW2}
  N.~Seiberg and E.~Witten,
  {\it Monopoles, duality and chiral symmetry breaking in N=2 supersymmetric QCD},
  \npb{431}{1994}{484},
  \hepth{9408099}.

\bibitem{Argyres}
  P.~C.~Argyres and M.~R.~Douglas,
  {\it New phenomena in SU(3) supersymmetric gauge theory,}
  \npb{448}{1995}{93},
  \hepth{9505062}.

\bibitem{Douglas}
  M.~R.~Douglas and S.~H.~Shenker,
  {\it Dynamics of SU(N) supersymmetric gauge theory,}
  \npb{447}{1995}{271},
  \hepth{9503163}.

\bibitem{PaikYaf}
  S.~Paik and L.~G.~Yaffe,
  {\it Thermodynamics of SU(2) $\Ntwo$ supersymmetric Yang-Mills theory},
  \jhep{1001}{2010}{059},
  \arXivid{0911.1392}.

\bibitem{Witten}
  E.~Witten,
  {\it Anti-de Sitter space, thermal phase transition, and confinement in gauge theories},
  \atmp{2}{1998}{505},
  \hepth{9803131}.

\bibitem{Csaki}
  C.~Csaki, H.~Ooguri, Y.~Oz and J.~Terning,
  {\it Glueball mass spectrum from supergravity},
  \jhep{9901}{1999}{017},
  \hepth{9806021}.

\bibitem{Koch}
  R.~de Mello Koch, A.~Jevicki, M.~Mihailescu and J.~P.~Nunes,
  {\it Evaluation of glueball masses from supergravity},
  \prd{58}{1998}{105009},
  \hepth{9806125}.

\bibitem{BMT}
  R.~C.~Brower, S.~D.~Mathur and C.~I.~Tan,
  {\it Glueball spectrum for QCD from AdS supergravity duality},
  \npb{587}{2000}{249},
  \hepth{0003115}.

\bibitem{KS}
  P.~K.~Kovtun and A.~O.~Starinets,
  {\it Quasinormal modes and holography},
  \prd{72}{2005}{086009},
  \hepth{0506184}.

\bibitem{DeWolfe}
  O.~DeWolfe and D.~Z.~Freedman,
  {\it Notes on fluctuations and correlation functions in holographic renormalization group flows},
  \hepth{0002226}.

\bibitem{deHaro}
  S.~de Haro, S.~N.~Solodukhin and K.~Skenderis,
  {\it Holographic reconstruction of spacetime and renormalization in the AdS/CFT correspondence},
  \cmp{217}{2001}{595},
  \hepth{0002230}.

\bibitem{Bianchi}
  M.~Bianchi, D.~Z.~Freedman and K.~Skenderis,
  {\it How to go with an RG flow},
  \jhep{0108}{2001}{041},
  \hepth{0105276}.

\bibitem{Aharony}
  O.~Aharony, A.~Buchel and P.~Kerner,
  {\it The black hole in the throat --- thermodynamics of strongly coupled cascading gauge theories},
  \prd{76}{2007}{086005},
  \arXivid{0706.1768}.

\bibitem{Recipes}
  W.~H.~Press, B.~P.~Flannery, S.~A.~Teukolsky, W.~T.~Vetterling,
  {\it Numerical Recipes in C: The Art of Scientific Computing},
  Cambridge Univ. Press, 2nd ed. (1992).




\end{thebibliography}
\end{document}